\documentclass[preprint,review, 12pt]{elsarticle}
\makeatletter
\def\ps@pprintTitle{%
 \let\@oddhead\@empty
 \let\@evenhead\@empty
 \def\@oddfoot{\centerline{\thepage}}%
 \let\@evenfoot\@oddfoot}
\makeatother

\usepackage{lineno,hyperref}
\usepackage{array}
\usepackage{longtable}
\usepackage{float}
\usepackage{etoolbox}
\patchcmd{\pprintMaketitle}
 {\ifvoid\absbox\else\unvbox\absbox\par\vskip10pt\fi}
 {\ifvoid\absbox\else\clearpage\unvbox\absbox\par\vskip30pt\fi}
 {}{}
\patchcmd{\pprintMaketitle}
 {\hrule\vskip12pt}
 {}
 {}{}
\patchcmd{\pprintMaketitle}
 {\hrule\vskip12pt}
 {}
 {}{}
\appto{\pprintMaketitle}{}
\usepackage{afterpage}
\usepackage[font=small]{caption}
\usepackage{subfloat}
\usepackage{graphicx}
\usepackage{subcaption}
\usepackage{amsmath,amsfonts}
\allowdisplaybreaks
\newcommand{\RomanNumeralCaps}[1]
    {\MakeUppercase{\romannumeral #1}}
\usepackage{caption}

\begin{document}

\begin{frontmatter}

\title{A Thermodynamic Framework for Additive Manufacturing, using Amorphous Polymers, Capable of Predicting Residual Stress, Warpage and Shrinkage}
\author{P Sreejith \fnref{a1}}
\author{K Kannan \fnref{a1}}
\author{K R Rajagopal \fnref{a2}}
\address[a1]{Department of Mechanical Engineering, Indian Institute of Technology Madras, Chennai 600036, India}
\address[a2]{Department of Mechanical Engineering, Texas A\&M University, College Station, TX 77804, US}
%==================================ABSTRACT==================================================
\begin{abstract}
 A thermodynamic framework has been developed for a class of amorphous polymers used in fused deposition modeling (FDM), in order to predict the residual stresses and the accompanying distortion of the geometry of the printed part (warping). When a polymeric melt is cooled, the inhomogeneous distribution of temperature causes spatially varying volumetric shrinkage resulting in the generation of residual stresses. Shrinkage is incorporated into the framework by introducing an isotropic volumetric expansion/contraction in the kinematics of the body. We show that the parameter for shrinkage also appears in the systematically derived rate-type constitutive relation for the stress. The solidification of the melt around the glass transition temperature is emulated by drastically increasing the viscosity of the melt.\\
\indent In order to illustrate the usefulness and efficacy of the derived constitutive relation, we consider four ribbons of polymeric melt stacked on each other such as those extruded using a flat nozzle: each layer laid instantaneously and allowed to cool for one second before another layer is laid on it. Each layer cools, shrinks and warps until a new layer is laid, at which time the heat from the newly laid layer flows and heats up the bottom layers. The residual stresses of the existing and newly laid layers readjust to satisfy equilibrium. Such mechanical and thermal interactions amongst layers result in a complex distribution of residual stresses. The plane strain approximation predicts nearly equibiaxial tensile stress conditions in the core region of the solidified part, implying that a preexisting crack in that region is likely to propagate and cause failure of the part during service. The free-end of the interface between the first and the second layer is subjected to the largest magnitude of combined shear and tension in the plane with a propensity for delamination.
\end{abstract}

\end{frontmatter}
%\linenumbers
%================================INTRODUCTION================================================

\section{Introduction}

\indent Manufacturing processes such as fused deposition modelling (FDM), extrusion molding, injection molding and blow molding induce residual stresses in the thermoplastic components. Determining these stresses is crucial, since it alters the mechanical response of the final solidified component.  Following instances demonstrate the same: (1) Plastic pipes manufactured by extrusion, develop compressive residual stresses on the outer surfaces and tensile residual stresses on the inner surfaces. Creep tests carried out to study crack propagation on v-notched specimens cutout from these plastic pipes, confirmed that compressive stresses hindered crack growth  whereas tensile stresses aided it (see Chaoui et. al. \cite{Chaoui}). (2) Similarly, quenching of  thermoplastics induce compressive residual stresses in the material. Three point bending tests and Izod impact tests, conducted to determine the fatigue life and impact strength of the quenched specimens, respectively, imply that, the strength increases with the magnitude of the residual stresses (see Hornberger et. al. \cite{Hornberger} and So et. al.  \cite{So}). \\
\indent In this paper, we attempt to develop a theory to capture the thermodynamic process associated with FDM during the layer by layer fabrication of a polymeric component. We aim to achieve the following: (1) Develop a thermodynamic framework that can represent the FDM process. (2) Use the developed model to: (a) Determine the residual stresses induced in the final solidified component due to rapid cooling and (b) capture the dimensional changes (shrinking and warping) of the part, which manifests as a consequence of the residual stresses.\\
\indent In recent years, FDM has been proving to be useful in the fabrication of patient specific transplants, tissue scaffolds and  pharmaceutical products (see Melchels et. al. \cite{MELCHELS}; Melocchi et. al. \cite{MELOCCHI}). The use of this technology in the medical industry has been growing fast due to the flexibility it offers in fashioning products, such as tools for surgical planning and medical implants. For instance, it has replaced solvent casting, fiber-bonding, membrane lamination, melt molding, and gas forming, which were previously used to fabricate medical grade polymethylmethacrylate (PMMA) implants for craniofacial reconstruction and regeneration surgeries (see Espanil et.al. \cite{Espalin2010}; Nyberg et. al. \cite{Nyberg}). One of the most important advantages of FDM is the possibility to architect the microstructure of the part. This flexibility was demonstrated by Muller et. al.  \cite{Muller}, wherein they fabricated an implant for a frontal-parietal defect with branching vascular channels to promote the growth of blood vessels.\\
\indent Stress-shielding effect caused by metal implants have led researchers to turn to polymers such as polyetheretherketone (PEEK) for orthopedic implants (see Han et.al. \cite{Han}). PEEK displays mechanical behaviour similar to that of the cortical bone (see Vaezi and Yang \cite{Vaezi} ), due to the presence of ether and ketone groups along the backbone of it's molecular structure. There have been surgeries such as cervical laminoplasty (see Syuhada et. al. \cite{Syuhada} ), total knee replacement surgery(see Swathi and Devadath \cite{Harish}  ; Borges et. al. \cite{BORGES} ), patient specific upper limb prostheses attachment (see Kate et. al. \cite{Kate} ) and patient specific tracheal implants (see Freitag et. al. \cite{Freitag}) recorded in the literature, which use PEEK, polylactic acid (PLA), acrylonitrile butadiene styrene (ABS), polyurethane (PU) and other medical grade polymer implants that have been fabricated using FDM. Recently, additive manufacturing has received a huge impetus in the healthcare sector due to the ongoing Covid-19 pandemic, and the technology is being used to fabricate products such as ventilator valves, face shields, etc.\\
\indent During fabrication of such components, thermoplastics are heated above the melting point ($\theta_m$) and then rapidly cooled below the glass transition temperature ($\theta_g$), which causes drastic temperature gradients. Cooling reduces configurational entropy of the polymer molecules, and the sharp temperature gradients induce differential volumetric shrinkage, consequently leading to the formation of residual stresses. As previously stated, these stresses can be either detrimental or favourable with regard to the mechanical response of the components. \\
\indent Process parameters such as layer thickness, orientation, raster angle, raster width, air gap, etc. affect the component's final mechanical response (see Sood et. al. \cite{Sood2010}). For instance, the anistropic nature of a part fabricated by FDM (see Song et. al. \cite{SONG}) can be controlled by choosing the most suitable raster angle for each individual layer during the process, in such a way that the component will survive under the required service conditions. This has been demonstrated by Casavola et. al. \cite{CASAVOLA}, wherein a rectangular ABS specimen with a raster angle of $\pm {45}^{\circ}$ produced the least residual stresses in the specimen. The optimal raster angle was determined by repeated experimentation, with a different choice of the angle in each experiment. Developing a consistent theory for capturing the thermodynamic process and using the resulting constitutive relations in simulations can help in developing standardized methods with optimized parameters for manufacturing the component.\\ 
\indent Many of the theories currently available, concentrate on the analysis of polymers which are at a temperature below $\theta_g$, the reasoning being that, internal reaction forces due to volume shrinkage are too low in the melt phase to give rise to appreciable residual stresses (see Xinhua et. al. \cite{Xinhua}; Wang et. al. \cite{Wang}; Park et. al. \cite{KeunPark}, Macedo et. al. \cite{Quelho}). Although such an argument seems reasonable, in a process like FDM, the mechanical and thermal histories of the polymer melt have a huge impact on the mechanical response of the final component.\\
\indent Several computational frameworks have been developed, wherein some have assumed incompressible viscous fluid models for the melt phase (see Xia and Lu et. al. \cite{FDM1}; Dabiri et. al. \cite{Dabiri}, Comminal et. al. \cite{COMMINAL}) and linear elastic models (see Xinhua et. al. \cite{Xinhua}; Wang et. al. \cite{Wang}; Macedo et. al. \cite{Quelho}; Moumen et. al. \cite{Moumen}; Casavola et. al. \cite{CASAVOLA2}) or elasto-plastic models (see Armillotta et. al. \cite{ARMILLOTTA}; Cattenone et. al. \cite{CAT} ) for the solid phase. However, it is to be noted that polymer melts are generally modelled as viscoelastic fluids (see Doi and Edwards \cite{Doi}). This is supported by the fact that, residual stresses in injection molded components have been predicted much more accurately by viscoelastic models than by  elastic or elasto-plastic models (see Kamal et. al. \cite{MRKamal} and Zoetelief et. al. \cite{WFZoetelief}). In this paper, the polymer melts are assumed to be amorphous, and we make an entropic assumption to simplify the analysis. Hence, the internal energy is solely a function of temperature and therefore, at a constant temperature, the free energy accumulates only due to the reducing configurational entropy of the molecules, such as during stretching. When the load is removed, the melt again assumes a high entropy state, which is also the stress free configuration of the melt.\\
\indent Once the amorphous melt has been laid on the substrate, it undergoes phase transformation due to cooling.
Numerous ``ad hoc" methods have been implemented by various authors to capture the phase change.
 One of the simplest methods used was to assume the melt to transform into a solid at a material point where the temperature reaches $\theta_m$  (see Xia and Lu et. al. \cite{FDM2}) or $\theta_g$ (see Macedo et. al. \cite{Quelho}, see Armillotta et. al. \cite{ARMILLOTTA};  Moumen et. al. \cite{Moumen} and Cattenone et. al. \cite{CAT}). Generally, polymer melts undergo phase transition through a range of temperature values. The heat transfer between a polymer melt and the surroundings is usually proportional to the temperature difference between them, and solidification is initiated at the material points where the germ nuclei are activated into a growth nuclei. An amorphous polymer melt normally transforms into an amorphous solid, unless the melt consists of specific germ nuclei which can cause crystallization. In the case of FDM, such germ nuclei can be induced in the melt while extruding the melt out of the nozzle at high shear rates combined with a low temperature (see Northcutt et. al. \cite{NORTHCUTT2018182}).\\
\indent  In the transition regime, polymer molecules exist in both, the melt, and the solid phases simultaneously, and usually the body is modelled as a constrained mixture (see Rao and Rajagopal \cite{Rao2000};  Rao and Rajagopal \cite{Rao2002}; Kannan et. al. \cite{Kannan2002}; Kannan and Rajagopal \cite{Kannan2005}; and Kannan and Rajagopal \cite{Raj2004}). While considering such a complex phase transition process, it is much more prudent to assume the primary thermodynamic functions to be dependent on the weight fraction of the phases. The weight fraction can be easily represented as a linear function of temperature (see Liu et. al. \cite{Liu}) or it can be determined by statistical methods (see Avrami et. al. \cite{Avrami}).\\ 
\indent Considering this work to be a preliminary step in developing a consistent theory for FDM, we assume an amorphous polymer melt to transform into a solid through a drastic increase in the viscosity at the glass transition temperature ($\theta_g$). In such a phase change process, the melt transforms completely into a solid when the temperature at all the material points fall below the glass transition value ($\theta_g$). This assumption limits the applicability of our theory to very particular processes having certain specific process parameters, such as:(1) The temperature of the nozzle is maintained above $\theta_m$ throughout. (2) The intial temperature of the melt is also kept above $\theta_m$. These predefined conditions ensure that the melt does not undergo shear induced crystallization.  
The proposed framework is based on the general theory for viscoelastic fluids developed by  Rajagopal and Srinivasa \cite{KR2000} and on the theory of the crystallization of polymer melts developed by Rao and Rajagopal \cite{Rao2000} \cite{RAO2001} \cite{Rao2002}.  The framework requires an appropriate assumption of the Helmholtz potential and the rate of entropy production, and the evolution of the ``natural configuration" of the body is determined by maximizing the rate of entropy production.

%=======================================METHODOLOGY==========================================

\section{Methodology}
\subsection{Kinematics}
\begin{figure}[!h]
\centering
\includegraphics[scale=0.33]{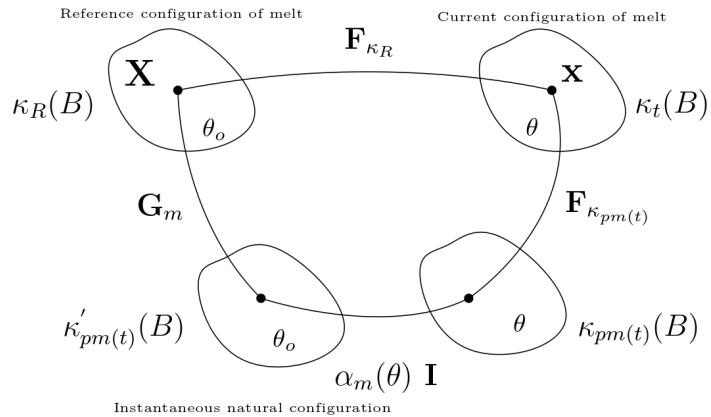}
\caption{ \sl The various configurations of the melt.\label{figure 3:}}
\end{figure}
 Let us consider the melt to be subjected to thermal loading. Initially the amorphous polymer melt (atactic polymer) is at the reference configuration $\kappa_R$(B) (refer to Fig.\ref{figure 3:}). This configuration is stress free and the polymer melt is at the reference temperature $\theta_{o}$. As the melt is cooled from $\theta_{o}$, it undergoes deformation and occupies the current configuration $\kappa_t $(B), at the current temperature $\theta$. The motion of the body from $\kappa _{R}$(B) to $\kappa _{t}$(B) is given by
\textbf{
\begin{equation}\label{mofun}
\textrm{x}=\textrm{$\chi$} _{\kappa_{R}} (\textrm{X},t).
\end{equation}}%
The motion is assumed to be a diffeomorphism. The velocity and the deformation gradient are given by
\textbf{
\begin{equation}\label{vel}
\textrm{v}=\frac{\partial \textrm{$\chi$} _{\kappa_{R}}}{\partial t},
\end{equation}}%
and
\textbf{
\begin{equation}\label{defgrad}
\textrm{F}_{\kappa_R} = \frac{\partial  \textrm{$\chi$} _{\kappa_R}}{\partial \textrm{X}}.
\end{equation}}%
 \indent The deformation gradient mapping $\kappa_{pm(t)}$(B) to $\kappa_t$(B) is $\textbf{F}_{\kappa_{pm(t)}}$. Let $\alpha_m(\theta) \textbf{I}$, where $\alpha_m(\theta)$ is a scalar valued funtion, represent pure thermal expansion/contraction. When $\alpha_m(\theta)=1$, i.e., an isothermal process, Fig.\ref{figure 3:} will coincide with the configurations employed by Rajagopal and Srinivasa \cite{KR2000}. The isothermally unloaded configuration $\kappa_{pm(t)}$(B) evolves due to thermal expansion/contraction of the melt (see Rajagopal and Srinivasa \cite{KR2004}). The configuration $\kappa^{'}_{pm(t)}$ is stress free, and since pure thermal expansion/contraction doesn't induce stresses in the material, the configuration $\kappa_{pm(t)}$ is also assumed to be stress free.
The deformation gradient is multiplicatively decomposed into a dissipative, thermal and an elastic part as shown
\textbf{
\begin{gather}\label{defmul}
\textrm{F}_{\kappa_{R}} = \alpha_m(\theta)\textrm{F}_{\kappa_{pm(t)}} \textrm{G}_m.
\end{gather}}%
In general, $\mathrm{\textbf{G}}_m$ need not be the gradient of a mapping. \\
\indent The right Cauchy-Green elastic stretch tensor and the left Cauchy-Green elastic stretch tensor are given by
\textbf{
\begin{gather}
%\begin{aligned}
\textrm{C}_{\kappa_R} = \textrm{F}_{\kappa_R}^{T} \textrm{F}_{\kappa_R},\hspace{0.5cm}\mathrm{and}\hspace{0.5cm}
\textrm{C}_{\kappa_{pm(t)}} = \textrm{F}_{\kappa_{pm(t)}}^{T} \textrm{F}_{\kappa_{pm(t)}}, \label{GS1}\\
\textrm{B}_{\kappa_R} = \textrm{F}_{\kappa_R} \textrm{F}_{\kappa_R}^{T},\hspace{0.5cm}\mathrm{and}\hspace{0.5cm}
\textrm{B}_{\kappa_{pm(t)}} = \textrm{F}_{\kappa_{pm(t)}} \textrm{F}_{\kappa_{pm(t)}}^{T}. \label{GS2}
%\end{aligned}
\end{gather}}%
The `right configurational stretch tensor' associated with the reference configuration ($\kappa_R$) and the instantaneous natural configuration ($\kappa^{'}_{pm(t)}$) is defined as
\textbf{
\begin{equation}\label{Conf_stretch}
\textrm{C}_{\kappa_R\rightarrow\kappa ^{'}_{pm(t)}}=\textrm{G}^{T}_m\textrm{G}_m.
\end{equation}}%
Therefore, Eq.$(\ref{GS2})_2$ can be represented as
\textbf{
\begin{equation}
\textrm{B}_{\kappa_{pm(t)}}=\frac{1}{\alpha^2_m(\theta)}\textrm{F}_{\kappa_R}\textrm{C}^{-1}_{\kappa_R\rightarrow\kappa^{'}_{pm(t)}}\textrm{F}^{T}_{\kappa_R}.\label{eq8}
\end{equation}}%
 The velocity gradients and the corresponding symmetric and skew-symmetric parts are defined as
\textbf{
\begin{equation}\label{velgrad1}
\textrm{L}=\dot{\textrm{F}}_{\kappa_{R}} \textrm{F}_{\kappa_{R}}^{-1},
\end{equation}}%
\textbf{
\begin{align}\label{velgradde1}
\textrm{D} & = \frac{\textrm{L} + \textrm{L}^{T}}{2}, & \textrm{W} & = \frac{\textrm{L} - \textrm{L}^{T}}{2},
\end{align}}
\textbf{
\begin{equation}\label{velgrad2}
\textrm{L}_{\kappa^{'}_{pm(t)}} =\dot{\textrm{G}}_m {\textrm{G}_m}^{-1},
\end{equation}}%
\textbf{
\begin{align}\label{velgradde2}
\textrm{D}_{\kappa^{'}_{pm(t)}} & = \frac{\textrm{L}_{\kappa^{'}_{pm(t)}} + \textrm{L}_{\kappa^{'}_{pm(t)}}^{T}}{2},&
\textrm{W}_{\kappa^{'}_{pm(t)}} & = \frac{\textrm{L}_{\kappa^{'}_{pm(t)}} - \textrm{L}_{\kappa^{'}_{pm(t)}}^{T}}{2}.
\end{align}}%
\indent The material time derivative of Eq.{(\ref{eq8})} will lead to
\textbf{
\begin{equation}\label{timeB}
\dot{\textrm{B}}_{\kappa_{pm(t)}}=-2\frac{\dot{\alpha}_m(\theta)}{\alpha_m(\theta)}\textrm{B}_{\kappa_{pm(t)}}+\textrm{L}\textrm{B}_{\kappa_{pm(t)}}+\textrm{B}_{\kappa_{pm(t)}}\textrm{L}^{T}+\frac{1}{\alpha^2_m(\theta)}\textrm{F}_{\kappa_R}\dot{\overline{\textrm{C}^{-1}}}_{\kappa_R\rightarrow\kappa^{'}_{pm(t)}}\textrm{F}^{T}_{\kappa_R}.
\end{equation}}%
The upper convected derivative is used to ensure frame invariance of the time derivatives, which, for any general second order tensor \textbf{A} is defined as
\textbf{
\begin{equation}\label{upper_convected}
    \overset{\bigtriangledown}{\textrm{A}}=\dot{\textrm{A}}-\textrm{L}\textrm{A}-\textrm{A}\textrm{L}^{T}.
\end{equation}}%
On comparing Eq.(\ref{timeB}) and Eq.(\ref{upper_convected}), we arrive at
\textbf{
\begin{equation}\label{time_upper}
\overset{\bigtriangledown}{\textrm{B}}_{\kappa_{pm(t)}}=-2\frac{\dot{\alpha}_m(\theta)}{\alpha_m(\theta)}\textrm{B}_{\kappa_{pm(t)}}+\frac{1}{\alpha^2_m(\theta)}\textrm{F}_{\kappa_R}\dot{\overline{\textrm{C}^{-1}}}_{\kappa_R\rightarrow\kappa^{'}_{pm(t)}}\textrm{F}^{T}_{\kappa_R}.
\end{equation}}%
Using Eq.(\ref{defmul}), Eq.(\ref{Conf_stretch}) and Eq.(\ref{time_upper}), we obtain
\textbf{
\begin{align}\label{evol1}
\overset{\bigtriangledown}{\textrm{B}} _{\kappa_{pm(t)}}= - 2\textrm{F}_{\kappa_{pm(t)}} \textrm{D}_{\kappa^{'}_{pm(t)}}\textrm{F}_{\kappa_{pm(t)}}^{T}-2\frac{\dot{\alpha}_m(\theta)}{\alpha_m(\theta)}\textrm{B}_{\kappa_{pm(t)}}.
\end{align}}%
 Note that, the evolution of $\textrm{\textbf{B}}_{\kappa_{pm(t)}}$ would depend on the structure of the constitutive equation chosen for $\alpha_m(\theta)$ (or more generally, on the material). In the event of an isothermal process, the scalar valued function $\alpha_m(\theta)$, will become unity, and Eq.(\ref{evol1}) will reduce to the form derived by Rajagopal and Srinivasa \cite{KR2000} as given below
\textbf{
\begin{align}\label{evol2}
\overset{\bigtriangledown}{\textrm{B}} _{\kappa_{pm(t)}}= - 2\textrm{F}_{\kappa_{pm(t)}} \textrm{D}_{\kappa^{'}_{pm(t)}}\textrm{F}_{\kappa_{pm(t)}}^{T}.
\end{align}}%
\indent To capture the effects of volume relaxation we will be using normalised invariants, where unimodular tensors are defined such that
\textbf{
\begin{gather}
\overset{-}{\textrm{C}}_{\kappa_{pm(t)}}=\mathrm{det}(\textrm{F}_{\kappa_{pm(t)}})^{-\frac{2}{3}}\textrm{C}_{\kappa_{pm(t)}}, \hspace{0.5cm}\mathrm{and}\hspace{0.5cm}
\overset{-}{\textrm{B}}_{\kappa_{pm(t)}}=\mathrm{det}(\textrm{F}_{\kappa_{pm(t)}})^{-\frac{2}{3}}\textrm{B}_{\kappa_{pm(t)}}. \label{nromin2}
\end{gather}}%
The invariants that we employ are defined through
\textbf{
\begin{gather} \label{nromin}
\textrm{\RomanNumeralCaps{1}}_{\kappa_{pm(t)}}=\mathrm{tr}(\overset{-}{\textrm{B}}_{\kappa_{pm(t)}}),\hspace{1cm} \textrm{\RomanNumeralCaps{2}}_{\kappa_{pm(t)}}=\mathrm{tr}(\overset{-}{\textrm{B}}_{\kappa_{pm(t)}})^{2},\hspace{0.2cm}\mathrm{and}\hspace{0.2cm} \textrm{\RomanNumeralCaps{3}}_{\kappa_{pm(t)}}=\mathrm{det}(\textrm{F}_{\kappa_{pm(t)}}).
\end{gather}}%
However, we plot Lode invariants (see Chen et. al. \cite{CHEN}) in section 3, to interpret the results better. This set of invariants are defined as
\begin{gather}\label{ortho_inv}
    \mathrm{K}_1 = \frac{\mathrm{tr}(\textbf{A})}{\sqrt{3}},\hspace{1cm}
    \mathrm{K}_2 = \sqrt{\mathrm{tr}({\textbf{A}}^{2}_d)},\hspace{0.2cm}\mathrm{and}\hspace{0.2cm}
    \mathrm{K}_3 = \frac{1}{3}\mathrm{sin}^{-1}\Big(\frac{\sqrt{6}\mathrm{tr}(\textbf{A}^{3}_d)}{\mathrm{tr}(\textbf{A}^{2}_d)^{\frac{3}{2}}}\Big),
\end{gather}
where $\textbf{A}$ is any tensor and $\textbf{A}_d$ is the deviatoric part of $\textbf{A}$. When $\mathrm{K}_1=0$, the tensor is purely composed of the deviatoric components. Similarly, when $\mathrm{K}_2=0$, the tensor is volumetric and $\mathrm{K}_3$ represents the mode of the tensor which varies from $\frac{-\pi}{6}$ to $\frac{\pi}{6}$.
%=========================================MODELLING==================================================

\subsection{Modeling}

%===================================MODELLING THE MELT======================================

\subsubsection{\textrm{Modeling the melt}}
\indent
  Helmholtz free energy per unit mass is defined with respect to the configuration $\kappa_{pm(t)}$(B), and is assumed to be a function of $\theta$ and the deformation gradient ${\textrm{F}}_{\kappa_{pm(t)}}$. We represent it as a sum of the contribution from the thermal interactions and the mechanical working, where the contribution due to mechanical working is assumed to be of the Neo-Hookean form. Further, we require it to be frame indifferent and isotropic with respect to the configuration $\kappa_{pm(t)}$(B), to arrive at
\begin{equation}\label{helmholtz}
\Psi_{m}(\theta,\textrm{\RomanNumeralCaps{1}}_{\kappa_{pm(t)}},\textrm{\RomanNumeralCaps{3}}_{\kappa_{pm(t)}})=\Psi^{th}_m(\theta)+\Psi^{mech}_m(\textrm{\RomanNumeralCaps{1}}_{\kappa_{pm(t)}},\textrm{\RomanNumeralCaps{3}}_{\kappa_{pm(t)}}),
\end{equation}
where
  \begin{equation}\label{helmholtz2}
\Psi^{th}_{m}(\theta)=\mathrm{A}^m+(\mathrm{B}^m+\mathrm{C}_{2}^m)(\theta-\theta_o)-\mathrm{C}_{1}^m\frac{(\theta-\theta_o)^2}{2}\\-\mathrm{C}_{2}^m\theta \mathrm{ln}\bigg(\frac{\theta}{\theta_o}\bigg),
\end{equation}
and
\textbf{
\begin{equation}\label{helmholtz3}
\Psi^{mech}_{m}(\textrm{\RomanNumeralCaps{1}}_{\kappa_{pm(t)}},\textrm{\RomanNumeralCaps{3}}_{\kappa_{pm(t)}})=\frac{\mu_1^m \theta}{2\rho_{{\kappa}_{pm(t)}} \theta_o}\Big[\mathrm{tr}(\overset{-}{\textrm{B}}_{\kappa_{pm(t)}})-3\Big]+\frac{k_1^m \theta}{2\rho_{{\kappa}_{pm(t)}} \theta_o}\Big[\mathrm{det}(\textrm{F}_{\kappa_{pm(t)}})-1\Big]^2,
\end{equation}}%
where $\theta_o, \hspace{0.1cm} \mathrm{A}^m,\hspace{0.1cm} \mathrm{B}^m,\hspace{0.1cm} \mathrm{C}_{1}^m,\hspace{0.1cm} \mathrm{C}_{2}^m,\hspace{0.1cm} \mu_{1}^m \hspace{0.1cm} \textrm{and} \hspace{0.1cm} k_{1}^m $ are the reference material temperature and the material constants respectively, and $ \mu_{1}^m, k_{1}^m \geq 0 $, and $\rho_{{\kappa}_{pm(t)}}$ is the density of the melt in the configuration $\kappa_{pm(t)}$(B).\\
\indent The rate of entropy production per unit volume due to mechanical working is defined as a function of $\theta$ and $\textbf{L}_{\kappa^{'}_{pm(t)}}$, and requiring the function to be frame indifferent and isotropic, we assume 
\textbf{
\begin{align} \label{dissipation}
\xi_m^{mech}(\theta,\textrm{D}_{\kappa^{'}_{pm(t)}})=\eta_1^m (\theta ){\mathrm{dev}(\textrm{D}_{\kappa^{'}_{pm(t)}}).\mathrm{dev}(\textrm{D}_{\kappa^{'}_{pm(t)}})}+\eta_2^m (\theta) \mathrm{tr}(\textrm{D}_{\kappa^{'}_{pm(t)}})^2,
\end{align}}%
where $\eta_1^m \hspace{0.1cm} \textrm{and}\hspace{0.1cm} \eta_2^m$ are the shear and bulk modulus of viscosity and $\eta_1^m , \eta_2^m \geq 0 $. The rate of mechanical dissipation should always be positive,
\textbf{
\begin{equation}\label{21}
\xi_m^{mech}\geq0,
\end{equation}}%
and the form given by Eq.(\ref{dissipation}), ensures that Eq.(\ref{21}) is always met.\\
\indent The balance of energy can be written as
\textbf{
\begin{equation}\label{balen}
\rho _{\kappa_t} \dot{\epsilon}_m = \textrm{T.D} - \mathrm{div}(\textrm{q}) + \rho _{\kappa_t} r,
\end{equation}}%
where \textbf{T} is the Cauchy stress tensor, $\dot{\epsilon}_m$ is the rate of change of internal energy per unit mass of the melt with respect to time, \textbf{q} is the heat flux and $r$ is the radiation of heat per unit mass of the melt. We express the second law of thermodynamics as an equality by introducing a rate of entropy production term, and it takes the form (see Green and Naghdi \cite{GREEN_NAGHDI})
\textbf{
\begin{equation}\label{ener-diss}
\textrm{T.D} - \rho _{\kappa_t} \dot{\Psi}_m - \rho _{\kappa_t} \eta_m \dot{\theta} - \frac{1}{\theta}\textrm{q}.\mathrm{grad}(\theta) = \rho _{\kappa_t} \theta \zeta_m=\xi^{tot}_m,\hspace{1cm}\xi^{tot}_m\geq0,
\end{equation}}%
where $\eta_m$ is the entropy per unit mass of the melt, $\theta$ is the absolute temperature, $\zeta_m$ is the rate of entropy production per unit mass of the melt and $\xi^{tot}_m$ is the total rate of entropy production per unit volume.
Substituting Eq.(\ref{helmholtz}) into Eq.(\ref{ener-diss}), we can re-write Eq.(\ref{ener-diss}) as
\textbf{
\begin{multline}\label{29}
\textrm{T.D}-\rho_{\kappa_t} \bigg \lbrace \frac{\partial \Psi^{th}_m}{\partial \theta}+ \frac{\mu_1^m }{2 \rho_{\kappa_{pm(t)}} \theta_o}\Big[\mathrm{tr}(\overset{-}{\textrm{B}}_{\kappa_{pm(t)}})-3\Big]+\frac{k_1^m}{2 \rho_{\kappa_{pm(t)}} \theta_o}\Big[\mathrm{det}(\textrm{F}_{\kappa_{pm(t)}})-1\Big]^2 + \eta_m \bigg \rbrace \dot{\theta}\\- \rho_{\kappa_t} \bigg \lbrace \frac{\mu_1^m \theta}{2 \rho_{\kappa_{pm(t)}} \theta_o}\mathrm{tr}(\dot{\overset{-}{\textrm{B}}}_{\kappa_{pm(t)}})-\frac{\mu_1^m \theta \dot{\rho}_{\kappa_{pm(t)}}}{2 \rho_{\kappa_{pm(t)}}^2 \theta_o}\Big[\mathrm{tr}(\overset{-}{\textrm{B}}_{\kappa_{pm(t)}})-3\Big]+\frac{k_1^m \theta}{ \rho_{\kappa_{pm(t)}} \theta_o}\Big[\mathrm{det}(\textrm{F}_{\kappa_{pm(t)}})-1\Big]\\ \dot{\overline{\mathrm{det}(\textrm{F}_{\kappa_{pm(t)}}})}-\frac{k_1^m \theta \dot{\rho}_{\kappa_{pm(t)}}}{2 \rho_{\kappa_{pm(t)}}^2 \theta_o}\Big[\mathrm{det}(\textrm{F}_{\kappa_{pm(t)}})-1\Big]^2 \bigg \rbrace - \frac{1}{\theta}\textrm{q}.\mathrm{grad}(\theta)=\xi^{tot}_m,\hspace{0.5cm}\xi^{tot}_m\geq0.
\end{multline}}%
\indent The balance of mass for the melt undergoing deformation between the configurations $\kappa_{R}$(B) and $\kappa_{pm(t)}$(B) is
\textbf{
\begin{gather}\label{balmass1}
\rho_{\kappa_{R}}=\rho_{\kappa_{pm(t)}} \mathrm{det}(\textrm{G}_m)(\alpha_m(\theta))^3,
\end{gather}}%
where $\rho_{\kappa_{R}}$ is the density at the reference configuration, $\kappa_{R}$(B). The balance of mass for the melt between the configurations $\kappa_{R}$(B) and $\kappa_{t}$(B) is given by
\textbf{
\begin{gather}\label{balmass2}
\rho_{\kappa_{R}}=\rho_{\kappa_t}\mathrm{det}(\textrm{F}_{\kappa_R}),
\end{gather}}%
where $\rho_{\kappa_{t}}$ is the density at the current configuration, $\kappa_{t}$(B).
Taking the material derivative of Eq.(\ref{balmass1}), we arrive at
\textbf{
\begin{gather}\label{densityder}
\frac{\dot{\rho}_{\kappa_{pm(t)}}}{\rho_{\kappa_{pm(t)}}}=-\mathrm{tr}(\textrm{D}_{\kappa^{'}_{pm(t)}})-\frac{3\dot{\alpha}_m(\theta)}{\alpha_m(\theta)}.
\end{gather}}%
Next, from Eq.$(\ref{GS2})_2$ and Eq.(\ref{evol1})
\textbf{
\begin{gather}
\dot{\overline{\mathrm{det}(\textrm{F}_{\kappa_{pm(t)}})}}=\mathrm{det}(\textrm{F}_{\kappa_{pm(t)}})\bigg \lbrace \mathrm{tr}(\textrm{D}-\textrm{D}_{\kappa^{'}_{pm(t)}})-3\frac{\dot{\alpha_m}}{\alpha_m} \bigg \rbrace, \label{derdetB}
\end{gather}}%
and from Eq.(\ref{evol1}), Eq.(\ref{nromin2}) and Eq.(\ref{derdetB}), we obtain
\textbf{
\begin{gather}
\mathrm{tr}(\dot{\overset{-}{\textrm{B}}}_{\kappa_{pm(t)}})=2 \mathrm{dev}(\overset{-}{\textrm{B}}_{\kappa_{pm(t)}}).(\textrm{D}-\textrm{D}_{\kappa^{'}_{pm(t)}})  \label{dertraceB}
\end{gather}}%
It follows from Eq.(\ref{defmul}) and  Eq.(\ref{29})-(\ref{derdetB}), that
\textbf{
\begin{multline}\label{subeq}
\bigg \lbrace \textrm{T}-\frac{ \mu_1^m \theta}{\mathrm{det}(\textrm{F}_{\kappa_{pm(t)}}) \theta_o}\mathrm{dev}(\overset{-}{\textrm{B}}_{\kappa_{pm(t)}})-\frac{k_1^m \theta }{ \theta_o}  \Big[\mathrm{det}(\textrm{F}_{\kappa_{pm(t)}})-1\Big]\textbf{I}\bigg \rbrace . \textrm{D}\\-\rho_{\kappa_t} \bigg \lbrace \frac{\partial \Psi^{th}_m}{\partial \theta}+ \frac{\mu_1^m }{2 \rho_{\kappa_{pm(t)}} \theta_o}\Big[\mathrm{tr}(\overset{-}{\textrm{B}}_{\kappa_{pm(t)}})-3\Big]+\frac{k_1^m}{2 \rho_{\kappa_{pm(t)}} \theta_o}\Big[\mathrm{det}(\textrm{F}_{\kappa_{pm(t)}})-1\Big]^2 + \eta_m \bigg \rbrace \dot{\theta}\\+\frac{ \mu_1^m \theta}{\mathrm{det}(\textrm{F}_{\kappa_{pm(t)}}) \theta_o}\mathrm{dev}(\overset{-}{\textrm{B}}_{\kappa_{pm(t)}}).\textrm{D}_{\kappa^{'}_{pm(t)}}+\bigg \lbrace\frac{k_1^m \theta }{ \theta_o}  \Big[\mathrm{det}(\textrm{F}_{\kappa_{pm(t)}})-1\Big]\\-\frac{ \mu_1^m \theta}{2 \mathrm{det}(\textrm{F}_{\kappa_{pm(t)}}) \theta_o}\Big[\mathrm{tr}(\overset{-}{\textrm{B}}_{\kappa_{pm(t)}})-3\Big]-\frac{k_1^m \theta }{2 \mathrm{det}(\textrm{F}_{\kappa_{pm(t)}}) \theta_o}\Big[\mathrm{det}(\textrm{F}_{\kappa_{pm(t)}})-1\Big]^2\bigg\rbrace\\\bigg[\mathrm{tr}(\textrm{D}_{\kappa^{'}_{pm(t)}})+\frac{3 \dot{\alpha}_m(\theta)}{\alpha_m(\theta)}\bigg] - \frac{1}{\theta}\textrm{q}.\mathrm{grad}(\theta)=\xi^{tot}_m, \hspace{0.5cm}\xi^{tot}_m\geq0.
\end{multline}}%
 One of the ways to satisfy Eq.(\ref{subeq}) is to assume that the three equations given below hold
\textbf{
\begin{gather}\label{entropy}
 \eta_m=- \frac{\partial \Psi^{th}_m}{\partial \theta}- \frac{\mu_1^m }{2 \rho_{\kappa_{pm(t)}} \theta_o}\Big[\mathrm{tr}(\overset{-}{\textrm{B}}_{\kappa_{pm(t)}})-3\Big]-\frac{k_1^m}{2 \rho_{\kappa_{pm(t)}} \theta_o}\Big[\mathrm{det}(\textrm{F}_{\kappa_{pm(t)}})-1\Big]^2 ,
\end{gather}}%
\textbf{
\begin{gather}\label{Stress}
\textrm{T}=\frac{ \mu_1^m \theta}{\mathrm{det}(\textrm{F}_{\kappa_{pm(t)}}) \theta_o}\mathrm{dev}(\overset{-}{\textrm{B}}_{\kappa_{pm(t)}})+\frac{k_1^m \theta }{ \theta_o}  \Big[\mathrm{det}(\textrm{F}_{\kappa_{pm(t)}})-1\Big]\textrm{I},
\end{gather}}%
and
\textbf{
\begin{multline}\label{entropythermal}
\xi^{tot}_m =\textrm{T}.\textrm{D}_{\kappa^{'}_{pm(t)}}-\Psi_m^{MECH} \textrm{I}.\textrm{D}_{\kappa^{'}_{pm(t)}}+\frac{3\dot{\alpha}_m(\theta)}{\alpha_m(\theta)} \bigg \lbrace - \frac{\mu_1^m \theta}{2 \theta_o \mathrm{det}(\textrm{F}_{\kappa_{pm(t)}})}\Big[\mathrm{tr}(\overset{-}{\textrm{B}}_{\kappa_{pm(t)}})-3\Big]\\+ \frac{k_1^m \theta}{2 \theta_o \mathrm{det}(\textrm{F}_{\kappa_{pm(t)}}) }\Big[ \big(\mathrm{det}(\textrm{F}_{\kappa_{pm(t)}})\big)^2 -1\Big] \bigg \rbrace-\frac{1}{\theta}\mathrm{\textbf{q}}.\mathrm{grad}(\theta), \hspace{0.5cm} \xi^{tot}_m\geq0,
\end{multline}}%
where
\textbf{
\begin{equation}\label{Psi}
\Psi_m^{MECH}= \rho_{\kappa_t}\Psi_m^{mech}. 
\end{equation}}%
 The Helmholtz free energy is given by
 \textbf{
 \begin{equation}\label{Legendre}
 \Psi_m = \epsilon_m - \theta \eta_m.
 \end{equation}}%
 From Eq.(\ref{helmholtz}) and Eq.(\ref{entropy}), the internal energy is defined as
\textbf{
\begin{gather}
 \epsilon_m=\mathrm{A}^m-\mathrm{B}^m\theta_o+\frac{1}{2}{\mathrm{C}_1}^m( \theta^2- \theta^2_o)+ \mathrm{C}_2^m(\theta-\theta_o),
\end{gather}}%
and the specific heat is
\textbf{
\begin{gather}
\frac{\partial \epsilon_m}{\partial \theta}={\mathrm{C}_1}^m \theta + \mathrm{C}_2^m.
\end{gather}}%
  \indent When the process is homothermal, Eq.(\ref{entropythermal}) will reduce to
\textbf{
\begin{equation}\label{entropymech}
\xi^{tot}_m =\textrm{T}.\textrm{D}_{\kappa^{'}_{pm(t)}}-\Psi_m^{MECH} \textrm{I}.\textrm{D}_{\kappa^{'}_{pm(t)}}, \hspace{0.5cm} \xi^{tot}_m\geq0.
\end{equation}}%
Therefore, in the current analysis, it seems prudent to suppose the following
\textbf{
\begin{equation}\label{dissi_mech}
\xi_m^{mech} = \textrm{T}.\textrm{D}_{\kappa^{'}_{pm(t)}}-\Psi_m^{MECH} \textrm{I}.\textrm{D}_{\kappa^{'}_{pm(t)}}.
\end{equation}}%
 Eq.(\ref{dissi_mech}) can be re-written as

\textbf{
\begin{equation} \label{constraint}
\xi_m^{mech}(\textrm{D}_{\kappa^{'}_{pm(t)}},\theta)=\xi_m^{mech}-\textrm{T}.\textrm{D}_{\kappa^{'}_{pm(t)}}+\Psi_m^{MECH} \textrm{I}.\textrm{D}_{\kappa^{'}_{pm(t)}}=0.
\end{equation}}%
Now, we assume the natural configuration (i.e. $\textbf{D}_{\kappa^{'}_{pm(t)}}$) to evolve in such a way that, for a fixed $\textbf{B}_{\kappa^{'}_{pm(t)}}$ and $\theta$ at each time instant, the rate of entropy production due to mechanical working is maximum, subject to the constraint in Eq.(\ref{constraint}) (see Rajagopal and Srinivasa \cite{K2004}). Therefore, to extremize $\xi_m^{mech}$ constrained by Eq.(\ref{constraint}), we write the augmented form as
\textbf{
\begin{gather}
\Phi_{aug} = \xi_m^{mech} + \lambda (\xi_m^{mech} - \textrm{T}.\textrm{D}_{\kappa^{'}_{pm(t)}}+\Psi_m^{MECH} \textrm{I}.\textrm{D}_{\kappa^{'}_{pm(t)}}),
\end{gather}}%
where $\lambda$ is the Lagrange multiplier. On taking the derivative with respect to $\textrm{\textbf{D}}_{\kappa^{'}_{pm(t)}}$, we obtain
\textbf{
\begin{gather}
\frac{\partial \Phi_{aug}}{\partial\textrm{D}_{\kappa^{'}_{pm(t)}}}=\frac{(1+\lambda)}{\lambda}\frac{\partial \xi_m^{mech}}{\partial \textrm{D}_{\kappa^{'}_{pm(t)}}}+(\Psi_m^{MECH} \textrm{I} - \textrm{T}) =0. \label{diffaug}
\end{gather}}%
At any time instant, assuming $\textbf{B}_{\kappa_{pm(t)}}$ and $\theta$ to be fixed, $\textbf{D}_{\kappa^{'}_{pm(t)}}$ will take only those values which will satisfy  Eq.(\ref{diffaug}). Substituting Eq.(\ref{dissipation}) into Eq.(\ref{diffaug}) and taking the scalar product with $\textbf{D}_{\kappa^{'}_{pm(t)}}$
\textbf{
\begin{multline} \label{diffaug1}
\frac{\partial \Phi_{aug}}{\partial \textrm{D}_{\kappa^{'}_{pm(t)}}}.\textrm{D}_{\kappa^{'}_{pm(t)}}=\frac{2(1+\lambda)}{\lambda}\bigg[\eta_1^m(\theta)\mathrm{dev}(\textrm{D}_{\kappa^{'}_{pm(t)}}).\textrm{D}_{\kappa^{'}_{pm(t)}}+\eta_2^m(\theta)[\mathrm{tr}(\textrm{D}_{\kappa^{'}_{pm(t)}})]^2\bigg]\\+\Big[\Psi_m^{MECH}\mathrm{tr}(\textrm{D}_{\kappa^{'}_{pm(t)}})  - \textrm{T}.\textrm{D}_{\kappa^{'}_{pm(t)}}\Big] =0.
\end{multline}}%
Comparing Eq.(\ref{diffaug1}) with Eq.(\ref{constraint}), we get $\frac{(1+\lambda)}{\lambda}=\frac{1}{2}$. Since the melt is assumed to be isotropic, we can assume that
\textbf{
\begin{gather} \label{iso}
\textrm{F}_{\kappa_{pm(t)}}=\textrm{V}_{\kappa_{pm(t)}},
\end{gather}}%
Also the eigenvectors of $\textbf{B}_{\kappa_{pm(t)}}$ and $\textbf{D}_{\kappa^{'}_{pm(t)}}$ are the same and hence the tensors commute (see Rajagopal and Srinivasa \cite{KR2000}). Therefore, from Eq.(\ref{evol1}), Eq.(\ref{diffaug}) and Eq.(\ref{iso}), we arrive at the form for the upper convected derivative of $\textbf{B}_{\kappa_{pm(t)}}$, which will determine the evolution of natural configuration of the melt, as
\textbf{
\begin{gather}\label{evolution}
\overset{\bigtriangledown}{\textrm{B}} _{\kappa_{pm(t)}}=2\Bigg(\bigg(\frac{\Psi^{MECH}_m-\frac{\mathrm{tr}(\textrm{T})}{3}}{3\eta^m_2(\theta)}\bigg)\textrm{I}-\frac{1}{\eta^m_1(\theta)}\mathrm{dev}(\textrm{T})-\frac{\dot{\alpha}_m(\theta)}{\alpha_m(\theta)}\textrm{I}\Bigg)\textrm{B}_{\kappa_{pm(t)}}.
\end{gather}}%

%============================ APPLICATION =====================================

\section{Application of the model}
\indent A prototypical problem to test the efficacy of the theory that has been developed is considered, i.e., we assume a particular geometry to be fashioned by FDM and use the constitutive equations to determine the residual stresses induced in the material during fabrication and the consequent dimensional instability of the geometry.\\
\subsection{Fused Deposition Modeling}
\indent In FDM, a polymer filament is heated above the melting temperature and extruded out through a nozzle attached to a robotic arm. A stereolithography (STL) file, containing information of the 3D geometry to be made, is then fed into a processing system that controls the robotic arm. The polymer melt is laid layer by layer through the nozzle with the help of the arm to  get  the  desired  3D  geometry, and the melt is allowed to cool. Polymers  used  in  this  process  are  usually thermoplastics like ABS (Acrylonitrile Butadiene Styrene), PLA (Polylactic Acid),  PS (Polystyrene),  PC  (Polycarbonates), PEEK (Polyetheretherketone), PMMA (Polymethylmethacrylate)  and  elastomers (see González-Henríquez et. al. \cite{POLYSTYRENE_PRINT}). Thermoplastics  are preferred over thermosets,  because the former's melt viscosity enables smooth extrusion through the nozzle, and at the same time helps to retain the shape once the melt is laid.  Complex part geometries can also be made with the help of support materials.\\
\indent In the current analysis, we assume each of the layers to be very long and be made of a single raster. Further, we assume a rectangular cross section for the layers, which finally adds up to each of the layers being in the shape of a very long ribbon. To lay such a layer, we need a nozzle with an appropriate geometry, such as a nozzle with a slot (see Loffer et. al. \cite{LOFFLER}) or a flat head nozzle which is capable of evening out the melt into the desired shape (see Kim et. al. \cite{FLAT_NOZ}). Four layers of the polymer melt are assumed to be laid one over the other consecutively, in the direction of the length of the ribbon. All four layers are assumed to have a sufficiently long rectangular cross-sections, which enables plane strain conditions to be enforced (refer to Fig.\ref{ri}).\\
\subsection{Governing equations}
\indent The constitutive form for the scalar valued function $\alpha_m(\theta)$, which represents the isotropic volume expansion/contraction, can be deduced from Eq.(\ref{defmul}). When $t>0$, taking the determinant of Eq.(\ref{defmul})
\textbf{
\begin{equation}\label{therex2}
\alpha_{m}(\theta)=\Bigg(\frac{\rho_{\kappa_R}}{\rho_{\kappa_t} \mathrm{det}(\textbf{F}_{\kappa_{pm(t)}})\mathrm{det}(\textbf{G}_m)}\Bigg)^{\frac{1}{3}}.
\end{equation}}%
Dilatometry experiments that are conducted to measure the specific volume changes are homothermal processes, and hence, the deformation gradients, $\textbf{G}_m$ and $\mathrm{\textbf{F}}_{\kappa_{pm(t)}}$, shown in Fig.\ref{figure 3:}, can be assumed to be identity transformations. Thus, density will become solely a function of temperature and the isotropic volume expansion/contraction can be represented as
\textbf{
\begin{equation}\label{therex3}
\alpha_{m}(\theta)=\Bigg(\frac{\rho_{\kappa_R}}{\rho_{\kappa_t} }\Bigg)^{\frac{1}{3}}.
\end{equation}}%
The density can be expressed in terms of specific volume, and therefore, $\alpha_m(\theta)$ is obtained as
\textbf{
\begin{equation}\label{therex4}
\alpha_{m}(\theta)=\Bigg(\frac{v(\theta)}{v(\theta_o)}\Bigg)^{\frac{1}{3}}=\Bigg(\frac{v}{v_o}\Bigg)^{\frac{1}{3}},
\end{equation}}%
where $v$ is the specific volume at the current temperature $\theta$ and $v_o$ is the specific volume at the reference temperature $\theta_o$ which can be represented by the Tait equation (see Haynes \cite{WM}). At time $t=0$ second, i.e. at reference temperature (temperature of the reference configuration) $\theta=\theta_o$\\
\begin{equation}\label{therex1}
\alpha_m (\theta_o) = 1.
\end{equation}%
The coefficient of volumetric thermal expansion/contraction, $\hat{\alpha}_{m}(\theta)$, is defined as
\textbf{
\begin{equation}\label{VolCo}
\hat{\alpha}_{m}(\theta)=\frac{1}{v}\bigg(\frac{\partial v}{\partial \theta}\bigg)=\rho_{\kappa_t}\bigg(\frac{\partial (\frac{1}{\rho_{\kappa_t}})}{\partial \theta}\bigg).
\end{equation}}%
 The relationship between isotropic volume expansion/contraction $\alpha_{m}(\theta)$ and the coefficient of volumetric thermal expansion/contraction $\hat{\alpha}_{m}(\theta)$ is obtained by taking the derivative of Eq.(\ref{therex4}) with respect to the current temperature, $\theta$, as
\textbf{
\begin{equation}\label{therCoe}
\hat{\alpha}_{m}(\theta)=\frac{3}{\alpha_{m}(\theta)}\frac{\partial \alpha_{m}(\theta)}{\partial \theta}.
\end{equation}}%
\indent As the layers cool, the viscosity of the polymer melt shoots up at glass transition temperature ($\theta_g$). The variation of the bulk and shear viscosities of the melt are defined as
\textbf{
\begin{gather}
\eta^{m}_n(\theta)=\eta^{s}_n\frac{\bigg(1-\mathrm{tanh} \big(a(\theta-\theta_g)\big)\bigg)}{2}+\eta^l_n\frac{\bigg(1+\mathrm{tanh} \big(a(\theta-\theta_g)\big)\bigg)}{2}, \hspace{0.5cm} n=1,2,
\end{gather}}%
where $\eta^m_1$ is the shear viscosity and $\eta^m_2$ is the bulk viscosity, $\eta^{s}_1$ and $\eta^{s}_2$ are the shear and the bulk viscosities below the glass transition temperature ($\theta_{g}$), $\eta^l_{{1}}$ and $\eta^l_{{2}}$ are the shear and the bulk viscosities above the glass transition temperature (${\theta}_g$) and $a$ is a constant. The shear viscosity and the bulk viscosity above the glass transition, i.e., $\eta^l_{{1}}$ and $\eta^l_{{2}}$, are defined as\\
\textbf{
\begin{gather}
\eta^{l}_n(\theta)=\eta^{o}_n\mathrm{e}^{\bigg(\mathrm{C}\big(\frac{1}{\theta}-\frac{1}{\theta_o}\big)\bigg)},\hspace{0.5cm} n=1,2,
\end{gather}}%
where $\eta^{o}_1$ is the shear viscosity at the melting temperature ($\theta_m$), $\eta^{o}_2$ is the bulk viscosity at the melting temperature ($\theta_m$) and C is a constant.\\
\indent The dependent variables are assumed to be a function of the current co-ordinates and current time. They are the displacement vector $\overset{\sim}{{\textbf{u}}}(\textbf{x},t)$, the current temperature $\overset{\sim}{\theta}(\textbf{x},t)$ and the left Cauchy-Green stretch tensor as defined in Eq.(\ref{GS2}$)_2$, i.e.,  $\overset{\sim}{\mathrm{\textbf{B}}}_{\kappa_{pm(t)}}(\textbf{x},t)$. The bases in the local and the global coordinate system are aligned with each other. Therefore, defining the dependent variables with respect to the local coordinate system is as good as defining it with respect to the global co-ordinate system. The position vector of a particle in the reference configuration is defined as
\begin{gather}\label{ref_Co}
    \textbf{X}=\textbf{x}(t)- \overset{\sim}{{\textbf{u}}}(\textbf{x},t).
\end{gather}
The velocity vector in the current configuration is derived by taking the total time derivative of Eq.(\ref{ref_Co})
\begin{gather}
    \overset{\sim}{\textbf{v}}(\textbf{x},t) = \Big(\mathrm{\textbf{I}}-\frac{\partial(\overset{\sim}{\textbf{u}}(\textbf{x},t))}{\partial \textbf{x}}\Big)^{-1}\frac{\partial \overset{\sim}{{\textbf{u}}}(\textbf{x},t)}{\partial t},
\end{gather}
and the velocity gradient is defined as
\begin{gather}\label{eq61}
    \overset{\sim}{\mathrm{\textbf{L}}}(\textbf{x}, t)=\frac{\partial \overset{\sim}{\textbf{v}}(\textbf{x},t)}{\partial \textbf{x}}.
\end{gather}
 The components of the displacement vector are 
\begin{gather}
   \big[ \overset{\sim}{{\textbf{u}}}\big]=\begin{bmatrix}{u}_x(x,y,t)\\{u}_y(x,y,t)\end{bmatrix},
\end{gather}
where $\big[ \overset{\sim}{{\textbf{u}}}\big]$ represents the component form of the displacement vector and, ${u}_x(x,y,t)$ and ${u}_y(x,y,t)$ are the components in the $x$ and $y$ directions respectively.
We represent the components of the velocity vector as given below
\begin{gather}
   \big[ \overset{\sim}{{\textbf{v}}}\big]=\begin{bmatrix}{v}_x(x,y,t)\\{v}_y(x,y,t)\end{bmatrix},
\end{gather}
where
\begin{gather}
     {v}_x = \frac{\frac{\partial u_{x}}{\partial t}\big(\frac{\partial u_y}{\partial y}-1\big)-\frac{\partial u_x}{\partial y}\frac{\partial u_y}{\partial t}}{\frac{\partial u_x}{\partial x}+\frac{\partial u_y}{\partial y}-\frac{\partial u_x}{\partial x}\frac{\partial u_y}{\partial y}+\frac{\partial u_x}{\partial y}\frac{\partial u_y}{\partial x}-1},
     \end{gather}
    and
     \begin{gather}
     {v}_y = \frac{\frac{\partial u_{y}}{\partial t}\big(\frac{\partial u_x}{\partial x}-1\big)-\frac{\partial u_x}{\partial t}\frac{\partial u_y}{\partial x}}{\frac{\partial u_x}{\partial x}+\frac{\partial u_y}{\partial y}-\frac{\partial u_x}{\partial x}\frac{\partial u_y}{\partial y}+\frac{\partial u_x}{\partial y}\frac{\partial u_y}{\partial x}-1}.
\end{gather}
Consequently, the component form of the velocity gradient follows from Eq.(\ref{eq61}) as
\begin{gather}
    \big[\overset{\sim}{\mathrm{\textbf{L}}}\big] = \begin{bmatrix} \frac{\partial v_x(x,y,t)}{\partial x} & \frac{\partial v_x(x,y,t)}{\partial y}&0\\\frac{\partial v_y(x,y,t)}{\partial x} & \frac{\partial v_y(x,y,t)}{\partial y}&0\\0&0&0\end{bmatrix}.
\end{gather}
We represent the components  of $\overset{\sim}{\mathrm{\textbf{B}}}_{\kappa_{pm(t)}}(\textbf{x},t)$ and the stress tensor $\mathrm{\textbf{T}}\Big(\overset{\sim}{\theta}, \overset{\sim}{\mathrm{\textbf{B}}}_{\kappa_{pm(t)}}\Big)$ as
\begin{equation}
    \big[\overset{\sim}{\mathrm{\textbf{B}}}_{\kappa_{pm(t)}}\big] = \begin{bmatrix} \mathrm{{B}}_{xx}(x,y,t) & \mathrm{{B}}_{xy}(x,y,t)&0\\\mathrm{{B}}_{xy}(x,y,t) & \mathrm{{B}}_{yy}(x,y,t)&0\\0&0&\mathrm{{B}}_{zz}(x,y,t)\end{bmatrix},
\end{equation}
and
\begin{equation}
    \big[\mathrm{\textbf{T}}\big] = \begin{bmatrix} \mathrm{{T}}_{xx}(x,y,t) & \mathrm{{T}}_{xy}(x,y,t)&0\\\mathrm{{T}}_{xy}(x,y,t) & \mathrm{{T}}_{yy}(x,y,t)&0\\0&0&\mathrm{{T}}_{zz}(x,y,t)\end{bmatrix},
\end{equation}
where $\mathrm{T}_{xx}(x,y,t)$, $\mathrm{T}_{yy}(x,y,t)$, $\mathrm{T}_{xy}(x,y,t)$ and $\mathrm{T}_{zz}(x,y,t)$ are derived by substituting ${\theta}(x,y,t)$ and the components of $\overset{\sim}{\mathrm{\textbf{B}}}_{\kappa_{pm(t)}}(\textbf{x},t)$ into Eq.(\ref{Stress}).\\
\indent There are seven coupled field equations that have to be solved for, i.e., two components of the linear momentum equation, the energy equation [Eq.(\ref{balen})] and four components of the evolution equation [Eq.(\ref{evolution})] represented in the Eulerian form. The linear momentum balance is given by
\begin{equation}\label{1}
    \frac{\partial \mathrm{{T}}_{xx}}{\partial x}+\frac{\partial \mathrm{{T}}_{xy}}{\partial y} = 0,
\end{equation}
and
\begin{equation}\label{2}
    \frac{\partial \mathrm{{T}}_{xy}}{\partial x}+\frac{\partial \mathrm{{T}}_{yy}}{\partial y} = 0.
\end{equation}
We assume radiation to be absent in the current analysis and hence the energy equation is written as
\begin{multline}\label{3}
     \mathrm{{T}}_{xx}\frac{\partial v_x}{\partial x}+\mathrm{{T}}_{yy}\frac{\partial v_y}{\partial y}+\mathrm{{T}}_{xy}\bigg(\frac{\partial v_x}{\partial y}+\frac{\partial v_y}{\partial x}\bigg)+k\Big(\frac{\partial^{2}{\theta}}{\partial x^{2}}+\frac{\partial^{2}{\theta}}{\partial y^{2}}\Big) \\= {\rho}_{\kappa_t}\frac{\partial\epsilon({\theta})}{\partial {\theta}}\Big(\frac{\partial {\theta}}{\partial t}+\frac{\partial {\theta}}{\partial x}v_x+\frac{\partial {\theta}}{\partial y}v_y\Big).
\end{multline}
The evolution equations are
\begin{multline}\label{4}
    \frac{\partial{\mathrm{{{B}}}}_{xx}}{\partial t}+\frac{\partial {\mathrm{{{B}}}}_{xx}}{\partial x}v_x+\frac{\partial {\mathrm{{{B}}}}_{xx}}{\partial y}v_y-2\bigg(\frac{\partial v_x}{\partial x}\mathrm{{B}}_{xx}+\frac{\partial v_x}{\partial y}\mathrm{{B}}_{xy}\bigg)=\frac{2}{3\eta^{m}_2(\theta)}\Bigg[\Psi^{MECH}_m\\-\frac{(\mathrm{T}_{xx}+\mathrm{T}_{yy}+\mathrm{T}_{zz})}{3}\Bigg]\mathrm{{B}}_{xx}-\frac{2}{\eta^{m}_1(\theta)}\Bigg[\Big(\frac{2\mathrm{T}_{xx}}{3}-\frac{\mathrm{T}_{yy}}{3}-\frac{\mathrm{T}_{zz}}{3}\Big)\mathrm{{B}}_{xx}+\mathrm{T}_{xy}\mathrm{{B}}_{xy}\Bigg]\\-\frac{2}{\alpha_m({\theta})}\frac{\partial \alpha_m({\theta})}{\partial {\theta}}\Big(\frac{\partial {\theta}}{\partial t}+\frac{\partial {\theta}}{\partial x}v_x+\frac{\partial {\theta}}{\partial y}v_y\Big)\mathrm{{B}}_{xx},
\end{multline}%
\begin{multline}\label{5}
    \frac{\partial{\mathrm{{{B}}}}_{yy}}{\partial t}+\frac{\partial {\mathrm{{{B}}}}_{yy}}{\partial x}v_x+\frac{\partial {\mathrm{{{B}}}}_{yy}}{\partial y}v_y-2\bigg(\frac{\partial v_y}{\partial x}\mathrm{{B}}_{xy}+\frac{\partial v_y}{\partial y}\mathrm{{B}}_{yy}\bigg)=\frac{2}{3\eta^{m}_2(\theta)}\Bigg[\Psi^{MECH}_m\\-\frac{(\mathrm{T}_{xx}+\mathrm{T}_{yy}+\mathrm{T}_{zz})}{3}\Bigg]\mathrm{{B}}_{yy}-\frac{2}{\eta^{m}_1(\theta)}\Bigg[\Big(\frac{-\mathrm{T}_{xx}}{3}+\frac{2\mathrm{T}_{yy}}{3}-\frac{\mathrm{T}_{zz}}{3}\Big)\mathrm{{B}}_{yy}+\mathrm{T}_{xy}\mathrm{{B}}_{xy}\Bigg]\\-\frac{2}{\alpha_m({\theta})}\frac{\partial \alpha_m({\theta})}{\partial {\theta}}\Big(\frac{\partial {\theta}}{\partial t}+\frac{\partial {\theta}}{\partial x}v_x+\frac{\partial {\theta}}{\partial y}v_y\Big)\mathrm{{B}}_{yy},
\end{multline}%
\begin{multline}\label{6}
    \frac{\partial{\mathrm{{{B}}}}_{zz}}{\partial t}+\frac{\partial {\mathrm{{{B}}}}_{zz}}{\partial x}v_x+\frac{\partial {\mathrm{{{B}}}}_{zz}}{\partial y}v_y=\frac{2}{3\eta^{m}_2(\theta)}\Bigg[\Psi^{MECH}_m-\frac{(\mathrm{T}_{xx}+\mathrm{T}_{yy}+\mathrm{T}_{zz})}{3}\Bigg]\mathrm{{B}}_{zz}\\-\frac{2}{\eta^{m}_1(\theta)}\Bigg[\Big(\frac{-\mathrm{T}_{xx}}{3}-\frac{\mathrm{T}_{yy}}{3}+\frac{2\mathrm{T}_{zz}}{3}\Big)\mathrm{{B}}_{zz}\Bigg]-\frac{2}{\alpha_m({\theta})}\frac{\partial \alpha_m({\theta})}{\partial {\theta}}\Big(\frac{\partial {\theta}}{\partial t}+\frac{\partial {\theta}}{\partial x}v_x+\frac{\partial {\theta}}{\partial y}v_y\Big)\mathrm{{B}}_{zz},
\end{multline}%
and
\begin{multline}\label{7}
    \frac{\partial{\mathrm{{{B}}}}_{xy}}{\partial t}+\frac{\partial {\mathrm{{{B}}}}_{xy}}{\partial x}v_x+\frac{\partial {\mathrm{{{B}}}}_{xy}}{\partial y}v_y-\bigg(\frac{\partial v_x}{\partial x}\mathrm{{B}}_{xy}+\frac{\partial v_x}{\partial y}\mathrm{{B}}_{yy}\bigg)-\bigg(\frac{\partial v_y}{\partial x}\mathrm{{B}}_{xx}+\frac{\partial v_y}{\partial y}\mathrm{{B}}_{xy}\bigg)=\\\frac{2}{3\eta^{m}_2(\theta)}\Bigg[\Psi^{MECH}_m-\frac{(\mathrm{T}_{xx}+\mathrm{T}_{yy}+\mathrm{T}_{zz})}{3}\Bigg]\mathrm{{B}}_{xy}-\frac{2}{\eta^{m}_1(\theta)}\Bigg[\Big(\frac{2\mathrm{T}_{xx}}{3}-\frac{\mathrm{T}_{yy}}{3}-\frac{\mathrm{T}_{zz}}{3}\Big)\mathrm{{B}}_{xy}\\+\mathrm{T}_{xy}\mathrm{{B}}_{yy}\Bigg]-\frac{2}{\alpha_m({\theta})}\frac{\partial \alpha_m({\theta})}{\partial {\theta}}\Big(\frac{\partial {\theta}}{\partial t}+\frac{\partial {\theta}}{\partial x}v_x+\frac{\partial {\theta}}{\partial y}v_y\Big)\mathrm{{B}}_{xy},
\end{multline}%
where $\Psi^{MECH}_m$ in Eq.(\ref{4}-\ref{7}) is derived by substituting ${\theta}(x,y,t)$ and the components of $\overset{\sim}{\mathrm{\textbf{B}}}_{\kappa_{pm(t)}}(\textbf{x},t)$ into Eq.(\ref{Psi}). We solve the seven equations for the seven unknowns: two components of the displacement field ($\overset{\sim}{{\textbf{u}}}(\textbf{x},t)$), the temperature field ($\overset{\sim}{{\theta}}(\textbf{x},t)$) and four components of the left Cauchy-Green stretch tensor ($\overset{\sim}{\mathrm{\textbf{B}}}_{\kappa_{pm(t)}}(\textbf{x},t)$).
\subsection{Material properties}
 \indent The polymer melt is assumed to be of Polystyrene (PS). High molecular weight atactic 
 PS does not undergo crystallization (see Chai et. al. \cite{POLYSTYRENE}), and hence is a good candidate for our analysis. The thermal and mechanical properties of PS are given in Table \ref{Table1}. The mechanical properties provided in Table \ref{Table1} are associated with the $\kappa_{pm(t)}$ configuration. However, in the literature, the properties are associated with the $\kappa_t$ configuration. Therefore, we have assumed the values to lie in a ballpark range of the actual values. Since the values of shear modulus and bulk modulus given in Table \ref{Table1} are towards the higher end at temperatures above $\theta_g$, it leads to a very small relaxation time ($\frac{\eta^o_1}{\mu^m_1}=10^{-6}$ sec) of the melt. However, this seems to be a reasonable approximation.
\begin{center}
\begin{longtable}{l l l l}
\caption{Properties of PS}
\label{Table1}\\
\hline
\textbf{Material Properties} & \textbf{Value} & \textbf{Unit} & \textbf{Ref.}\\
\hline
\\
\textbf{Thermal properties}\\
Melting temperature, ${\theta}_m$ & $513$ & $\mathrm{K}$ & \cite{GLASS_MELTING} \\

Glass transition temperature, ${\theta}_g$ & $373$ & $\mathrm{K}$ &\cite{GLASS_MELTING} \\

Coefficient of heat \\conduction, $k$ (averaged value) & \vspace{0.1cm}$0.159$ & $\frac{\mathrm{W}}{\mathrm{mK}}$ & \cite{THERM_COND}\\

Coefficient of heat transfer\\ from the material to the& $90$ & $\frac{\mathrm{W}}{\mathrm{m}^2\mathrm{K}}$ & \cite{H}\\ environment, $h_e$ \\

Coefficient of heat transfer\\ from the material to the & $100$ & $\frac{\mathrm{W}}{\mathrm{m}^2\mathrm{K}}$ & \cite{H}\\ substrate, $h_s$ \\

Specific heat ($C_p$):\\
(Table 1a and 1b of \cite{Specific_heat})\\
$\mathrm{C}^{m}_1$ & 4.101  & $\frac{\mathrm{J}}{\mathrm{K}^2\mathrm{kg}}$ & \cite{Specific_heat}\\

 $\mathrm{C}^{m}_2$ &  11.16 & $\frac{\mathrm{J}}{\mathrm{Kkg}}$ & \cite{Specific_heat}\\
 
 Constants in the Tait equation:\\ 

$\mathrm{A}_3$ & 0.9287 & $\frac{\mathrm{cm}^{3}}{\mathrm{g}}$ & \cite{WM}\\

$\mathrm{A}_4$ & $5.131\mathrm{x}10^{-4}$ & $^oC^{-1}$ & \cite{WM}\\
\\
\textbf{Mechanical properties}\\

Shear modulus  ($\mu^{m}_1$) & $10^9$ & Pa & \cite{JEM} \\

Bulk modulus ($k^{m}_1$) & $3\mathrm{x}10^{10}$ & Pa & NA\\

Shear viscosity below glass transition\\ temperature, $\theta_{g}$ ($\eta^{s}_1$) & $10^{19}$ & Pas & NA\\

Bulk viscosity below glass transition\\ temperature, $\theta_{g}$ ($\eta^{s}_2$) & $2\mathrm{x}10^{20}$ & Pas & NA\\

Shear viscosity at the reference\\ temperature, $\theta_{o}$ ($\eta^{o}_1$) & $10^3$ & Pas & \cite{VISCOSITY}\\

Bulk viscosity at the reference\\ temperature, $\theta_{o}$ ($\eta^{o}_2$) & $2\mathrm{x}10^4$ & Pas & NA\\

 C & 22873 & $\mathrm{K}$ & \cite{VISCOSITY}\\
 
 $a$ & 0.7 & NA & NA\\

\hline

\end{longtable}
\end{center}%
\subsection{Initial and boundary conditions}
\indent The first layer of the melt is laid instantaneously on the substrate (which is assumed to be at the ambient temperature $\theta_{sub}=27^o$C throughout the process) at time t = 0 second. The surface that is in contact with the substrate is fixed along with the substrate acting as a heat sink. Free convection and traction free conditions are assumed on the outer surfaces (refer to Fig.\ref{rt1}). The layer is then allowed to cool for one second, at the end of which, the second layer is laid instantaneously on top of the first layer. Once the second layer is laid, continuity is established between the layers, thus both the layers act as a single contiguous body (refer to Fig.\ref{rt2}), i.e., we assume a perfect interface between the layers. The third and fourth layers are laid similarly one after the other with an interval of one second each  (refer to Fig.\ref{rt3} and  
\begin{figure}[H]
\centering
\vspace{3cm}
        \begin{subfigure}{0.9\textwidth}
                \includegraphics[width=12.5cm, height=6cm,keepaspectratio]{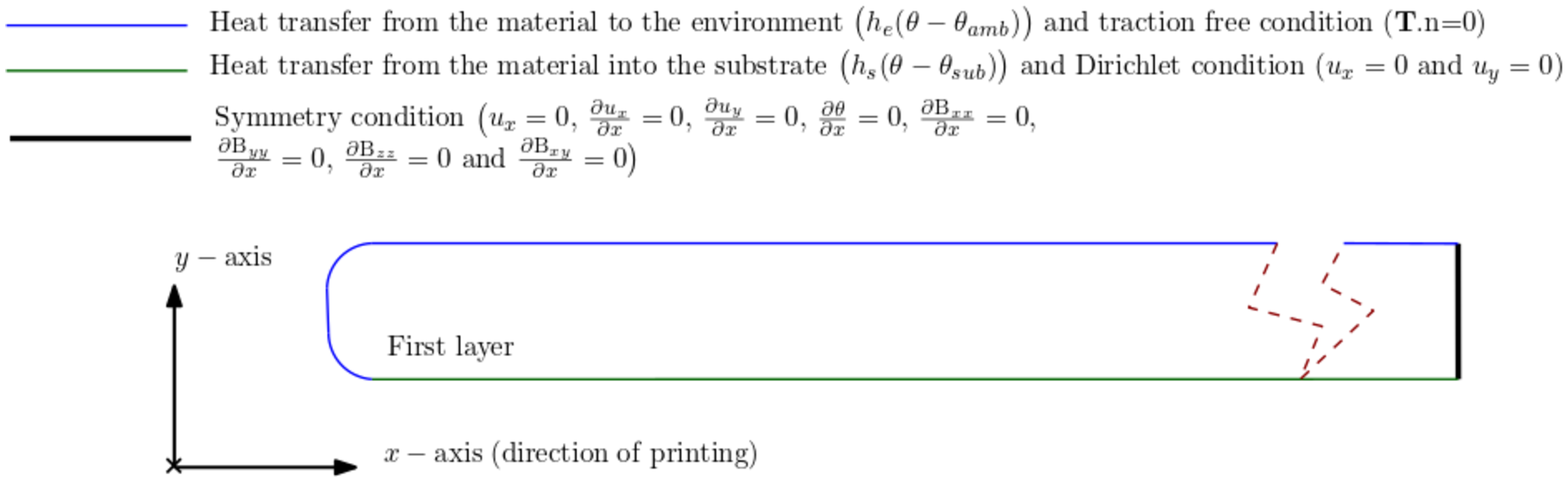}
                \caption{ \scriptsize The first layer laid at time t = 0 second and cooled till time t = 1 second.}
                \label{rt1}
        \end{subfigure}%
\vspace{3.5cm}
        \begin{subfigure}{0.9\textwidth}
                \includegraphics[width=12.5cm, height=6cm,keepaspectratio]{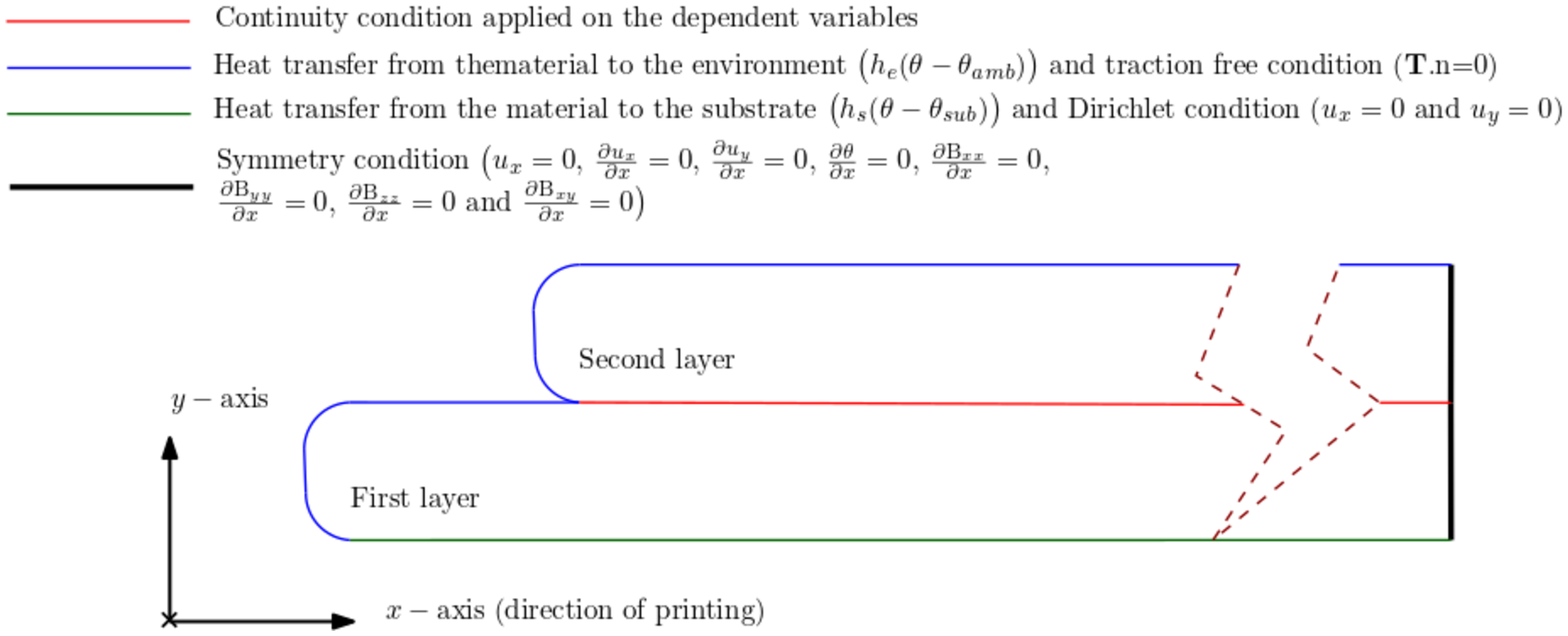}
                \caption{\scriptsize The second layer laid on top of the first layer at time t = 1 second and both layers cooled together till time t = 2 second.}
                \label{rt2}
        \end{subfigure}%
    \vspace{1cm}
 \end{figure}
\begin{figure}\ContinuedFloat
\centering
       
        \begin{subfigure}{0.9\textwidth}
         \includegraphics[width=12.5cm, height=6.5cm]{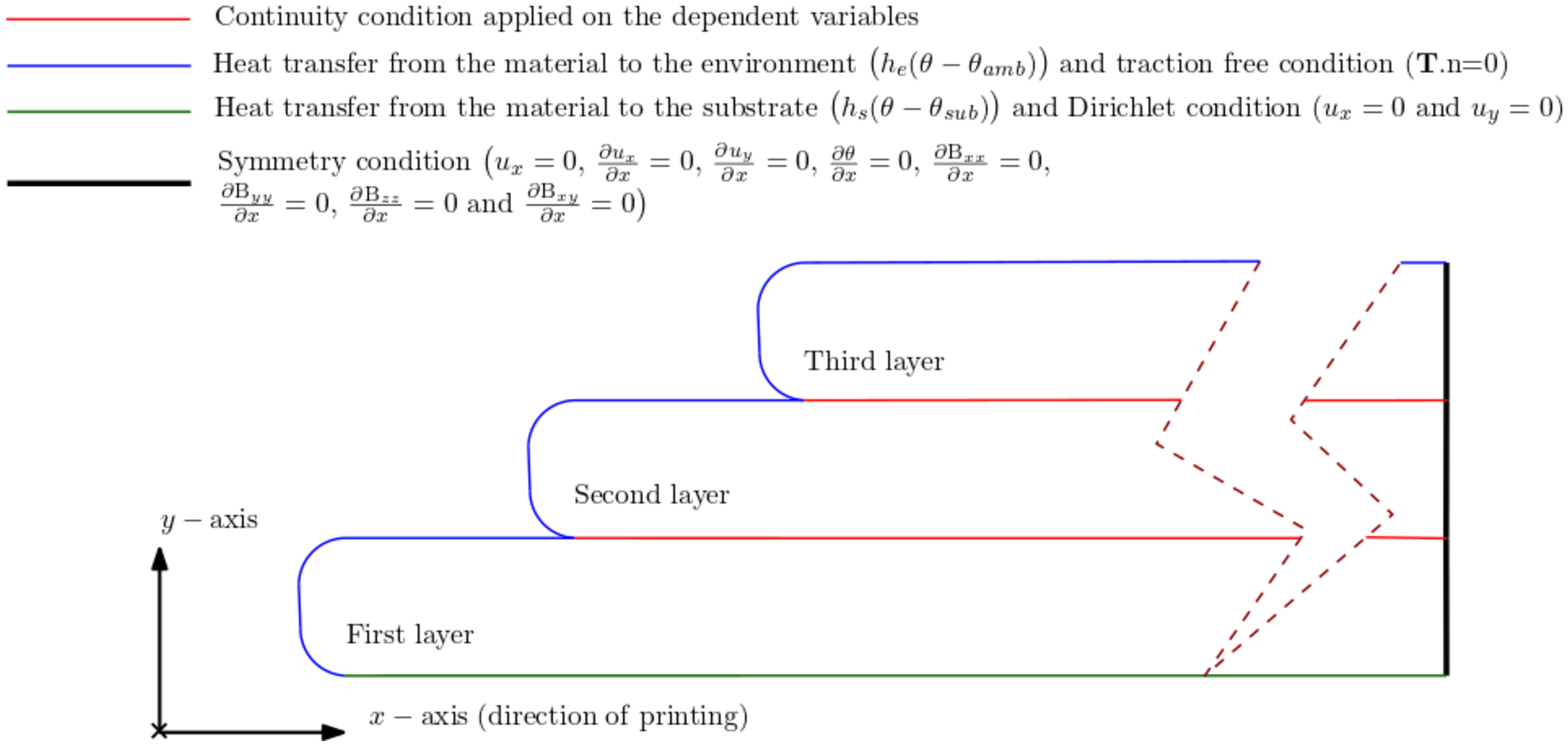}
                \caption{\scriptsize The third layer laid on top of the second layer at time t = 2 second and all the three layers cooled together till time t = 3 second.}
                \label{rt3}
        \end{subfigure}%
        \vspace{0.5cm}

        \begin{subfigure}{0.9\textwidth}
                \includegraphics[width=12.5cm, height=7cm]{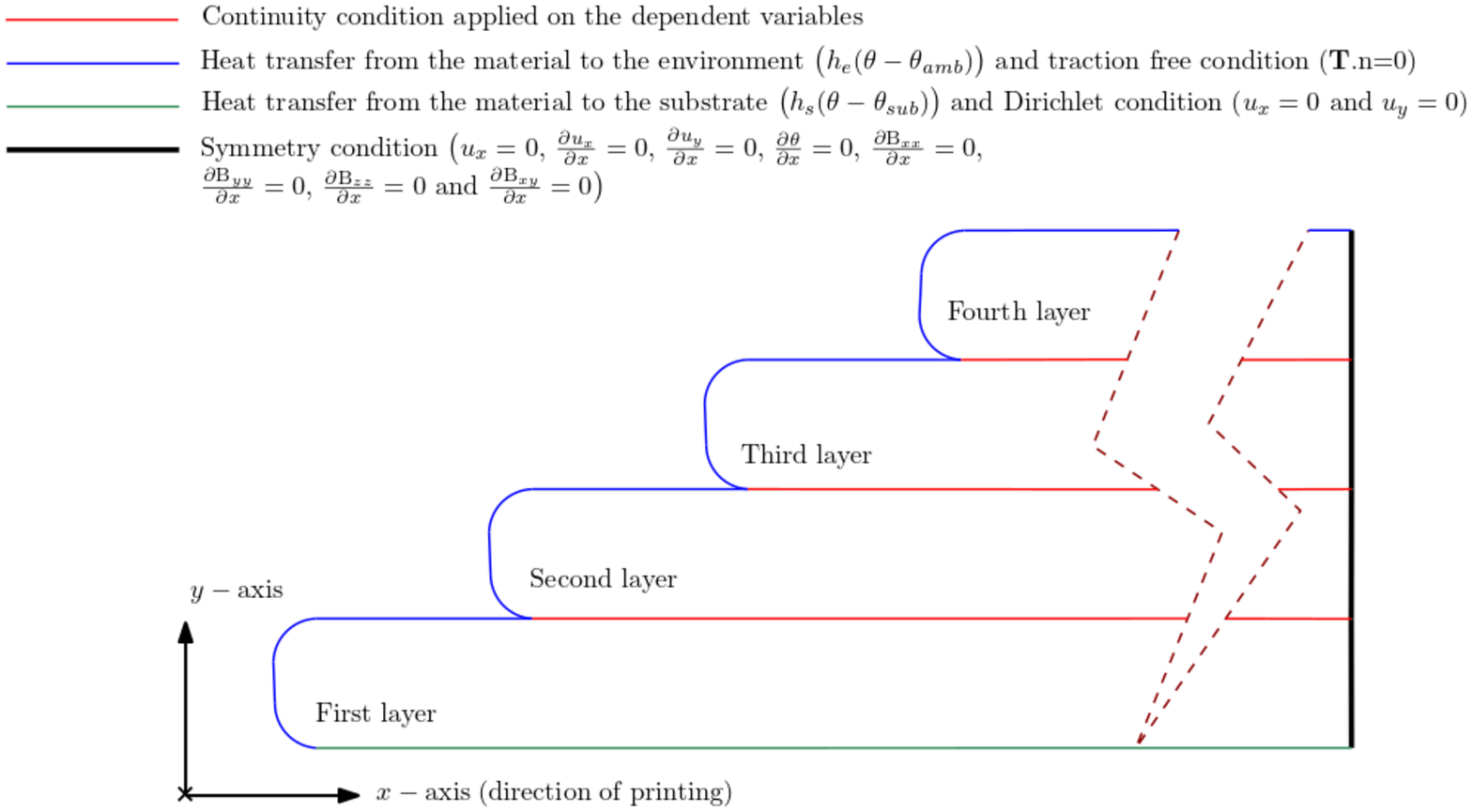}
                \caption{\scriptsize The fourth layer laid on top of the third layer at time t = 3 second and all the four layers cooled together till time t = 60 second.}
                \label{rt4}
        \end{subfigure}
        
        \caption{Representative images showing the boundary conditions on the layers at different time steps of the simulation. The dashed (dark red) line provides a break and is intended to denote that the layers are very long.}\label{ri}
\end{figure}%
\noindent Fig.\ref{rt4}).\\
\indent After all the layers have been laid, the entire system cools till the temperature field  attains a value close to the ambient temperature (preferably $27^o$C). Symmetry conditions have been imposed to reduce the computational time  (refer to Fig.\ref{rt1}, Fig.\ref{rt2}, Fig.\ref{rt3} and Fig.\ref{rt4}). The first, second, third and fourth layers have an aspect ratio of 80:1,76:1, 72:1 and 68:1 respectively.\\
\indent Before each layer is laid, the layer is at $241^o$C, which is $1^o$C above the melting point of Polystyrene (PS). At this temperature, the isotropic volume contraction/expansion ($\alpha_m(\overset{\sim}{\theta})$) is unity along with $\overset{\sim}{\textbf{B}}_{\kappa_{pm(t)}}(\textbf{x},t)$ being identity, i.e., 
\begin{equation}
    \big[\alpha_m(\theta_o)\mathrm{\textbf{I}}\big]=\big[\mathrm{\overset{\sim}{\textbf{B}}}_{\kappa_{pm(t)}}\big] = \begin{bmatrix} 1 & 0&0\\0 & 1&0\\0&0&1\end{bmatrix}\hspace{0.5cm}\textrm{at\hspace{0.1cm}t\hspace{0.1cm} =\hspace{0.1cm} 0\hspace{0.1cm} second}
\end{equation}
\subsection{Implementation}
\indent Representing non-linear equations in the Lagrangian description and integrating them in the material frame can cause excessive mesh distortions. This discrepancy of the method is because, each of the nodes of the mesh, follow the corresponding material points during the motion. The Eulerian description in which the nodes remain fixed while the material distorts, overcomes the above discrepancy. However, it compromises the numerical accuracy of the solution while computing with coarse meshes. It also fails to precisely identify the interface between domains, leading to poor numerical accuracy of the interfacial fluxes. \\
\indent The disadvantages of both the descriptions can be avoided by using the Arbitrary Lagrangian-Eulerian (ALE) method. ALE combines the best features of both Lagrangian and Eulerian descriptions. Each node of the ALE mesh is free to either remain fixed or rezone itself (see Donea et. al. \cite{ALE}), thus avoiding excessive mesh distortions and at the same time predicting the domain interfaces with better accuracy. It also provides comparatively better numerical accuracy of the solutions than the Eulerian approach.\\ 
\indent The non-linear equations, Eq.(\ref{1})-Eq.(\ref{7}), have been solved using ``Coefficient Form PDE" module in COMSOL $\mathrm{Multiphysics}^{{\mathrm{\scriptsize{TM}}}}$. The coefficients of the general PDE in the module, are populated by extracting the respective coefficients of the governing equations by using a MATLAB code. It is to be noted that the equations are provided to the module with respect to the spatial frame and are integrated in time by using the BDF method. The order of the scheme varies from 1 to 5 with free time stepping.
\subsection{Results and discussion}
\subsubsection{Volumetric shrinkage and residual stresses}
Temperature in the melt starts reducing as heat flows out through the free surfaces into the surroundings and through the fixed surface into the substrate (note that the rate of heat flow is higher into the substrate) and consequently $\alpha_m(\theta)$ begins to fall. Due to the rapid cooling process, the temperature reaches the ambient value in about 60 seconds. The temperature distribution in the layers  is given in Fig.\ref{temp} and the value of temperature ($\theta$) is higher towards the core than towards the surface, as is to be expected.\\
\indent Volumetric shrinkage in the first three seconds is not too evident. This can be attributed to the re-heating of the relatively cooler layer each time continuity is established between consecutive layers. Once all the layers are laid, and the entire system starts cooling, the volumetric shrinkage becomes
\begin{figure}[H]
\centering
\vspace{1cm}
        \begin{subfigure}{1\textwidth}
                \includegraphics[width=14cm, height=6cm]{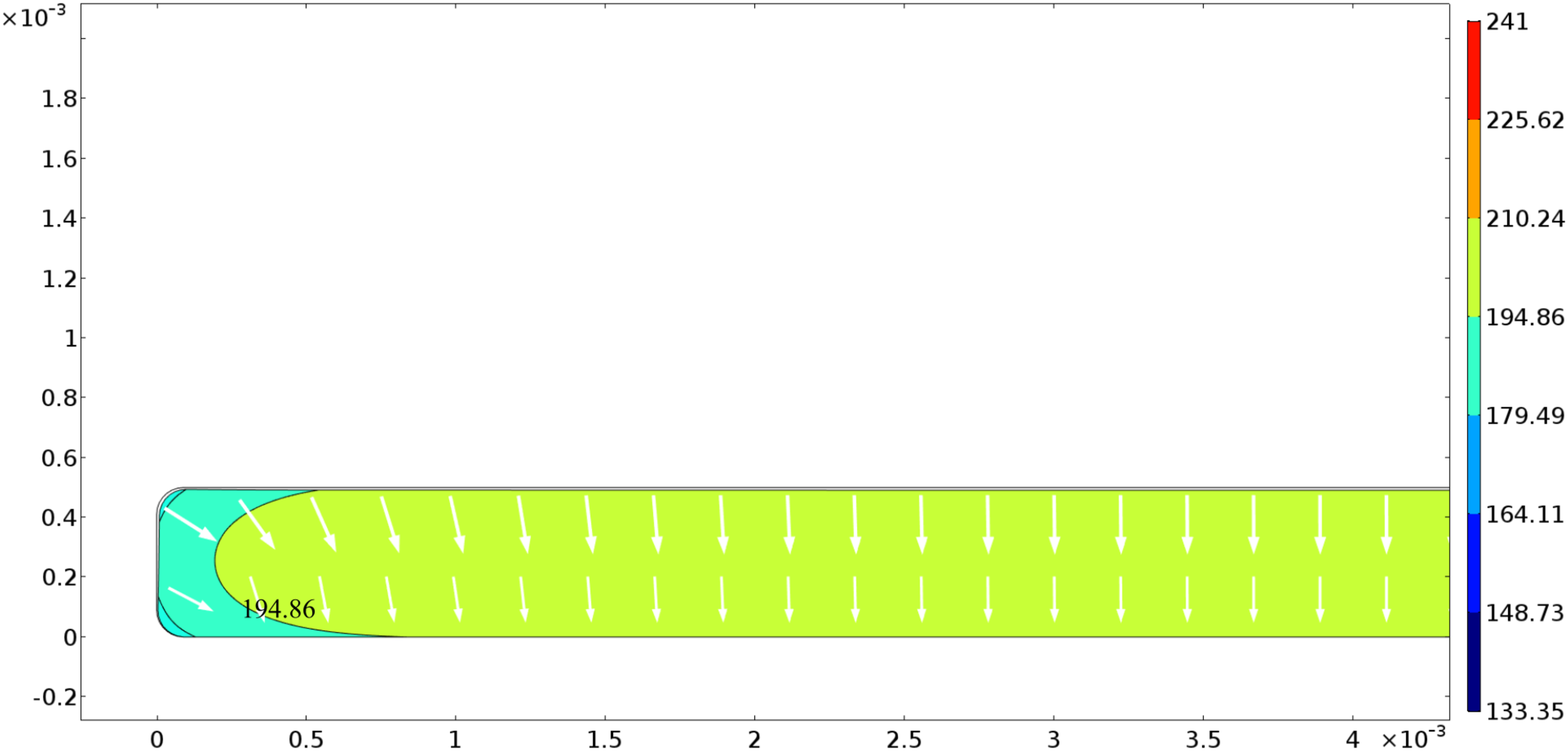}
                \caption{ \scriptsize Distribution of temperature in $^{o}\mathrm{C}$ at 1 second. The layer has cooled from the intial temperature of 241$^{o}\mathrm{C}$.}
                \label{t1}
        \end{subfigure}%
        \vspace{2cm}
        \begin{subfigure}{1\textwidth}
                \includegraphics[width=14cm, height=6cm]{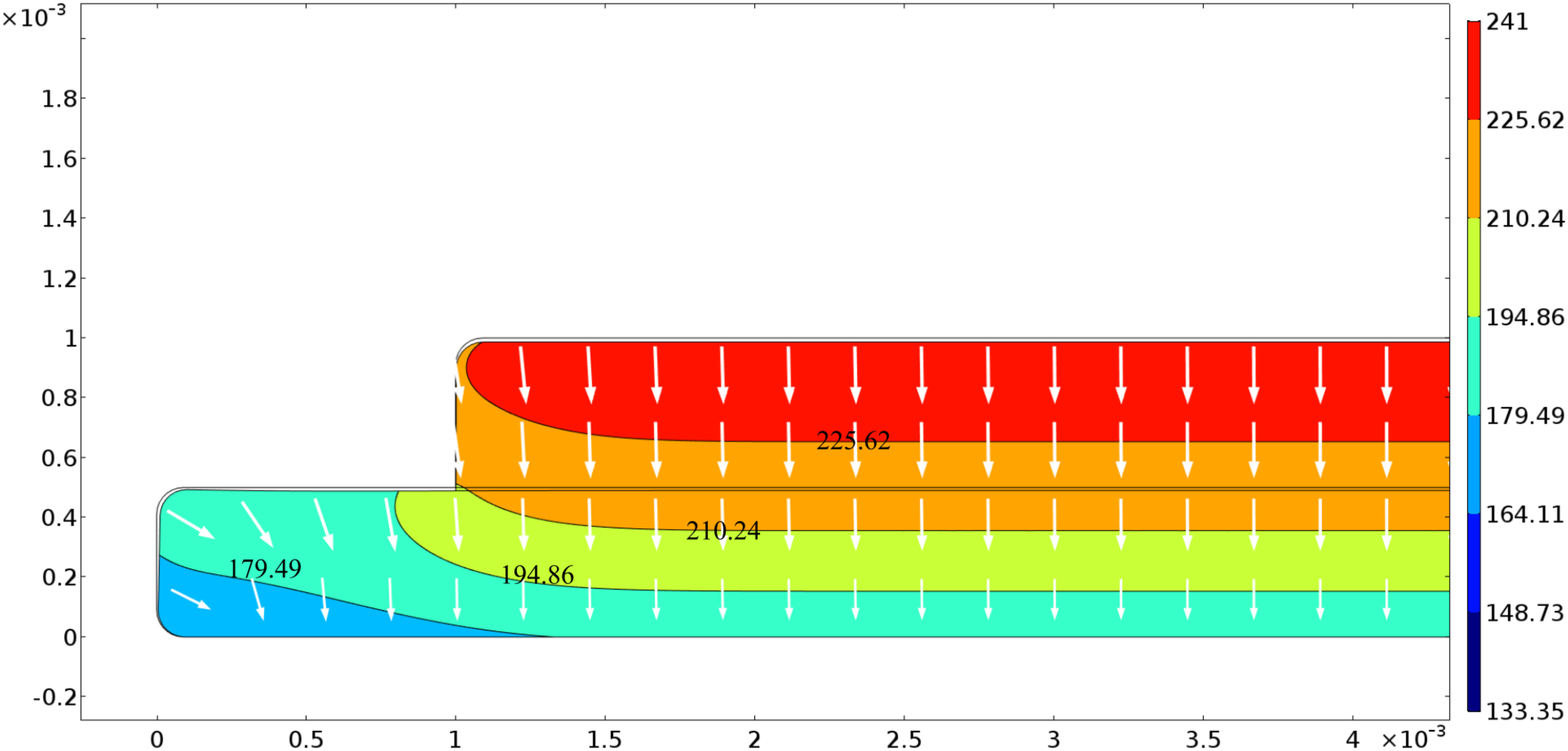}
                \caption{\scriptsize Distribution of temperature in $^{o}\mathrm{C}$ at 2 second. An increase in the temperature levels of the first layer is observed, which is attributed to the re-heating of the layer by the comparatively hotter second layer. }
                \label{t2}
        \end{subfigure}%
\end{figure}
\begin{figure}\ContinuedFloat
\centering 
        \begin{subfigure}{1\textwidth}
                \includegraphics[width=14cm,height=6cm]{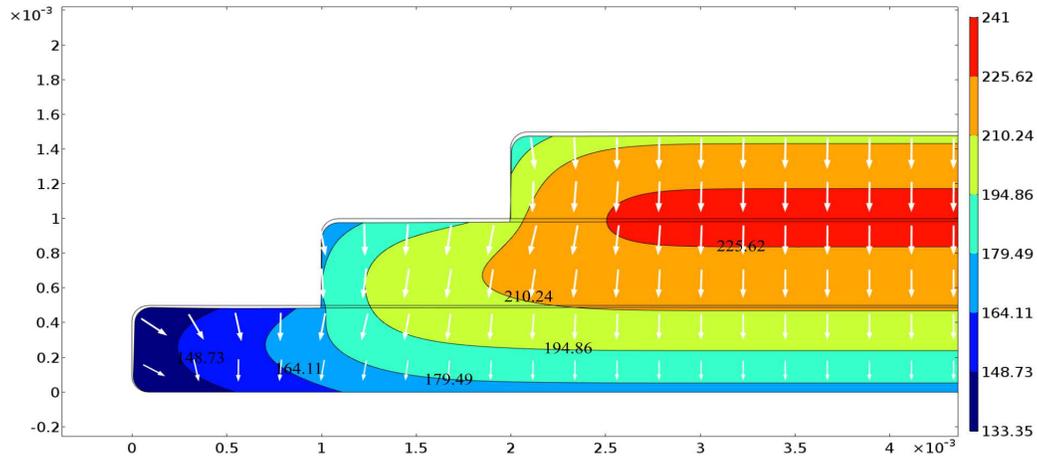}
                \caption{\scriptsize  Distribution of temperature in $^{o}\mathrm{C}$ at 3 second. The temperature levels in the second layer increases due to re-heating by the comparatively hotter third layer.}
                \label{t3}
        \end{subfigure}%
\vspace{0.5cm}     
        \begin{subfigure}{1\textwidth}
                \includegraphics[width=14cm, height=6cm]{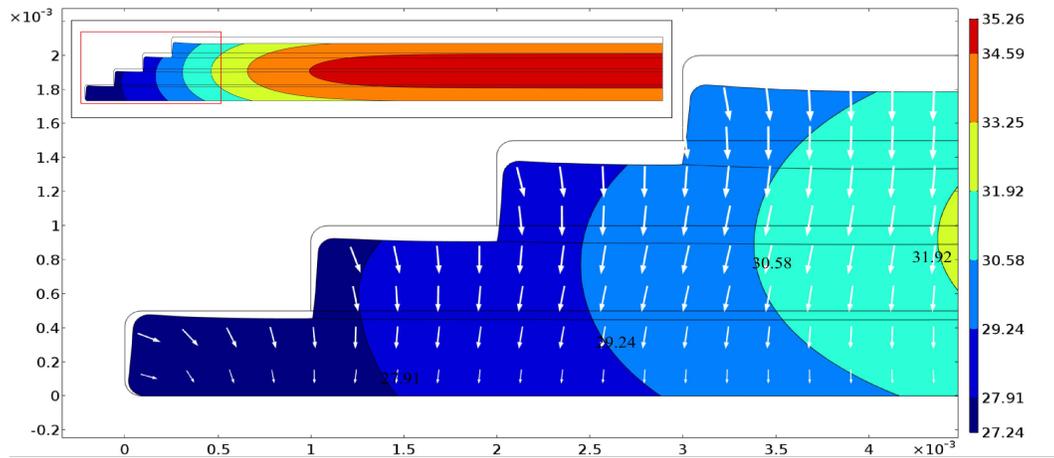}
                \caption{\scriptsize Distribution of temperature in $^{o}\mathrm{C}$ at 60 second. The inset shows the longest layer having a length of 0.02m, and the red box highlights the area of interest. The solid black line superimposed on the deformed configuration represents the initial geometry of the layers before warping and volumetric shrinkage has set in.}
                \label{t4}
        \end{subfigure}%
        \caption{Banded contour plots of temperature in $^{o}\mathrm{C}$, visualised on the deformed configuration. The white arrows depict the local displacement vector, and its length represents the magnitude. The relative length of arrows are in logarithmic scale. }\label{temp}
\end{figure}%
 \begin{figure}[H]
\centering
\vspace{1cm}
        \begin{subfigure}{1\textwidth}
                \includegraphics[width=14cm, height=6cm]{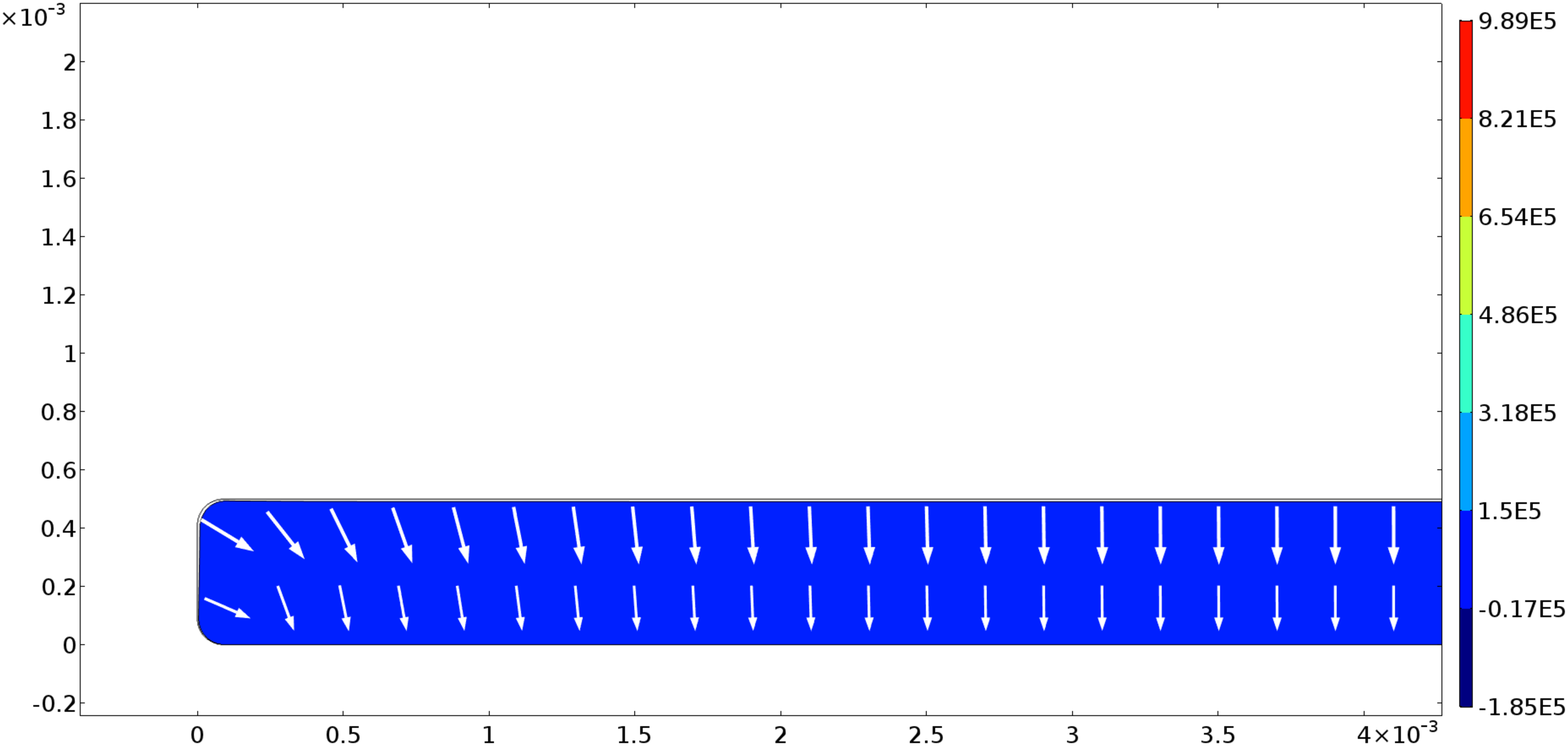}
                \caption{ \scriptsize Distribution of the mean normal stress (Pa), $\frac{\mathrm{K}_1}{\sqrt{3}}$, at 1 second. The stress values are low because the temperature values are well above $\theta_g$.}
                \label{VOL1}
        \end{subfigure}%
\vspace{2cm}
        \begin{subfigure}{1\textwidth}
                \includegraphics[width=14cm, height=6cm]{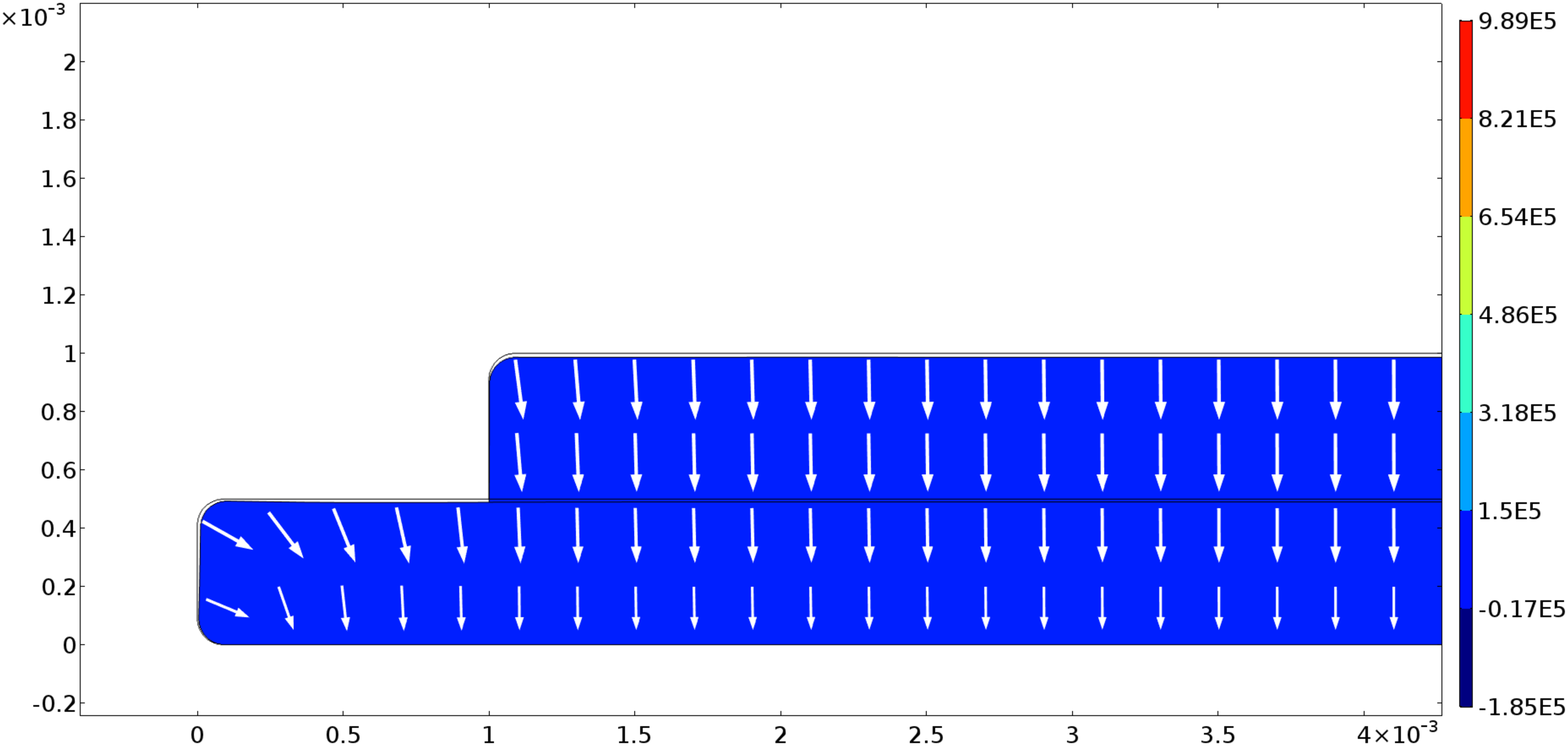}
                \caption{\scriptsize Distribution of the mean normal stress (Pa), $\frac{\mathrm{K}_1}{\sqrt{3}}$, at 2 second. Re-heating of the first layer ensures a delayed accumulation of stress in the material.}
                \label{VOL2}
        \end{subfigure}%
        \vspace{1cm}
\end{figure}
\begin{figure}\ContinuedFloat
\centering 
        \begin{subfigure}{1\textwidth}
                \includegraphics[width=14cm, height=6cm]{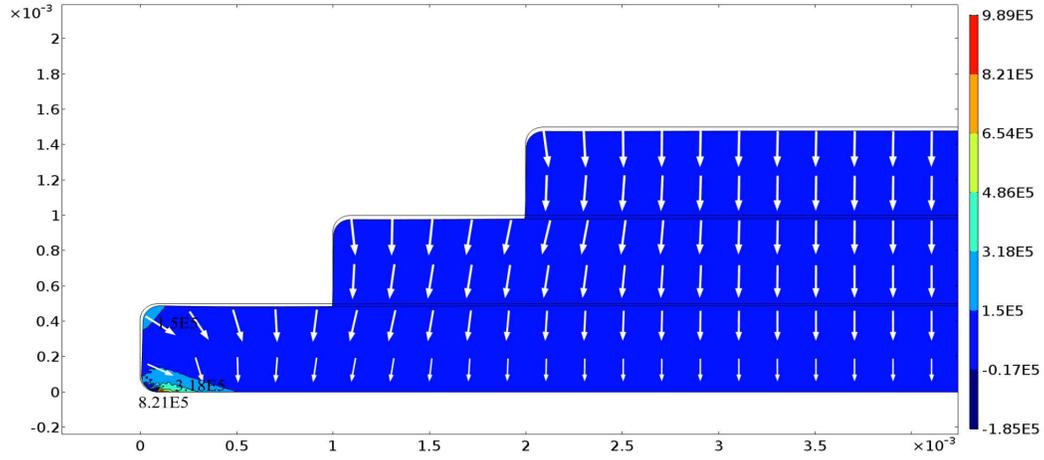}
                \caption{\scriptsize Distribution of the mean normal stress (Pa), $\frac{\mathrm{K}_1}{\sqrt{3}}$, at 3 second. Appreciable stress values start developing in the region where the temperature values are close to $\theta_g$. Re-heating delays the development of any significant stress at the core of the geometry.}
                \label{VOL3}
        \end{subfigure}%
\vspace{1cm}
         \begin{subfigure}{1\textwidth}
            \includegraphics[width=14cm, height=6cm]{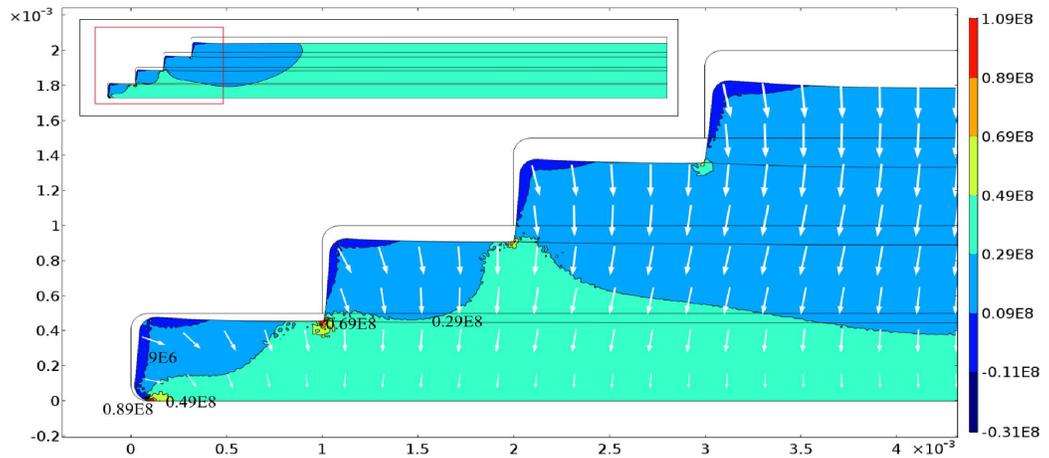}
                \caption{ \scriptsize Distribution of the mean normal stress (Pa), $\frac{\mathrm{K}_1}{\sqrt{3}}$, at 60 second. The stress values have shot up significantly, and it is higher towards the core, in comparison to the values at the regions close to the stepped end. }
                \label{VOL4}
        \end{subfigure}
        
        \caption{Banded contour plots of the mean normal stress (Pa), $\frac{\mathrm{K}_1}{\sqrt{3}}$, visualised on the deformed configuration.}\label{VOL}
\end{figure}%
 \begin{figure}[H]
\centering
\vspace{1cm}
        \begin{subfigure}{1\textwidth}
                \includegraphics[width=14cm, height=6cm]{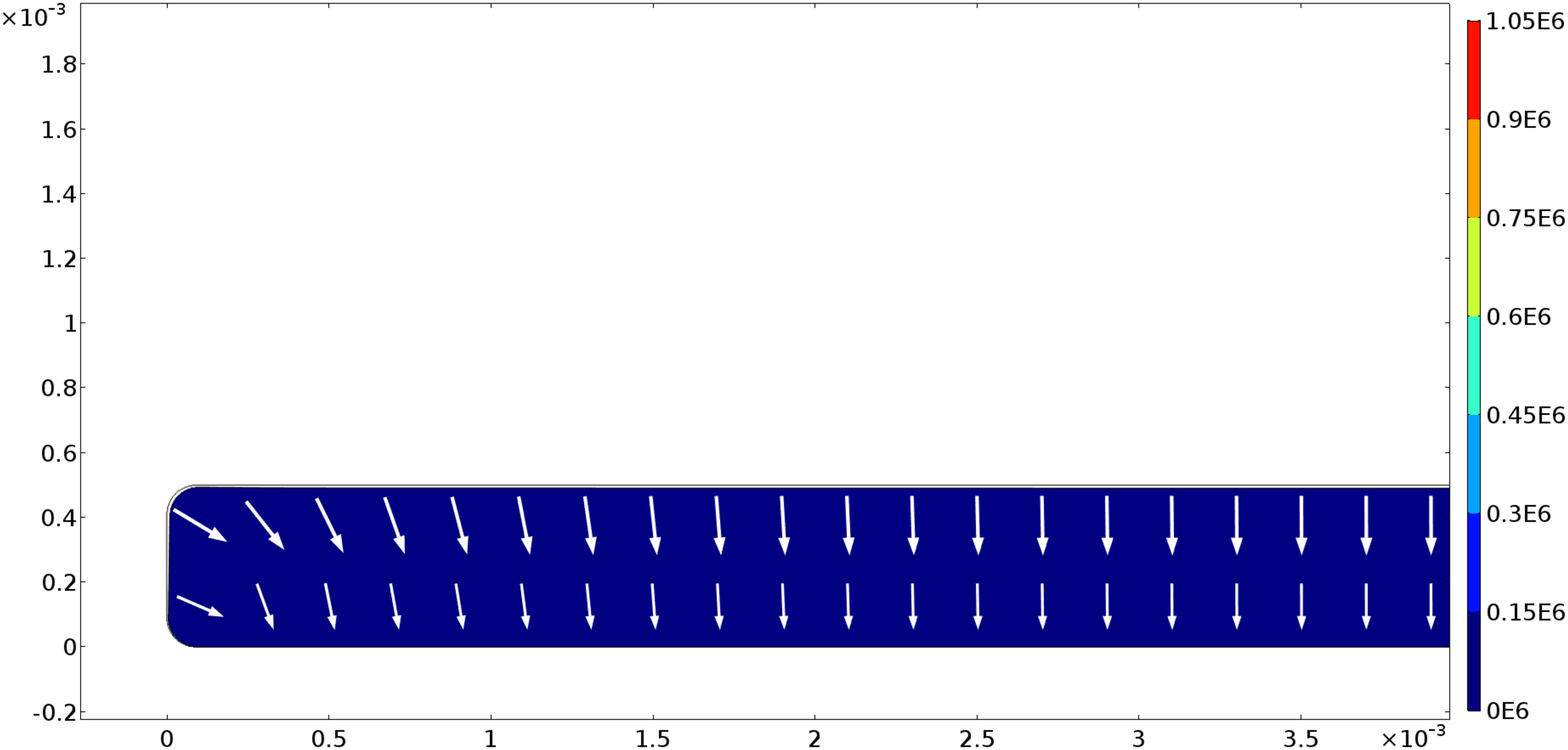}
                \caption{ \scriptsize Distribution of the norm of deviatoric part of stress (Pa), $\mathrm{K}_2$, at 1 second. The deviatoric stress values are also insignificant due to the temperature values being sufficiently above $\theta_g$.}
                \label{eivc1}
        \end{subfigure}%
        \vspace{2cm}
        \begin{subfigure}{1\textwidth}
                \includegraphics[width=14cm, height=6cm]{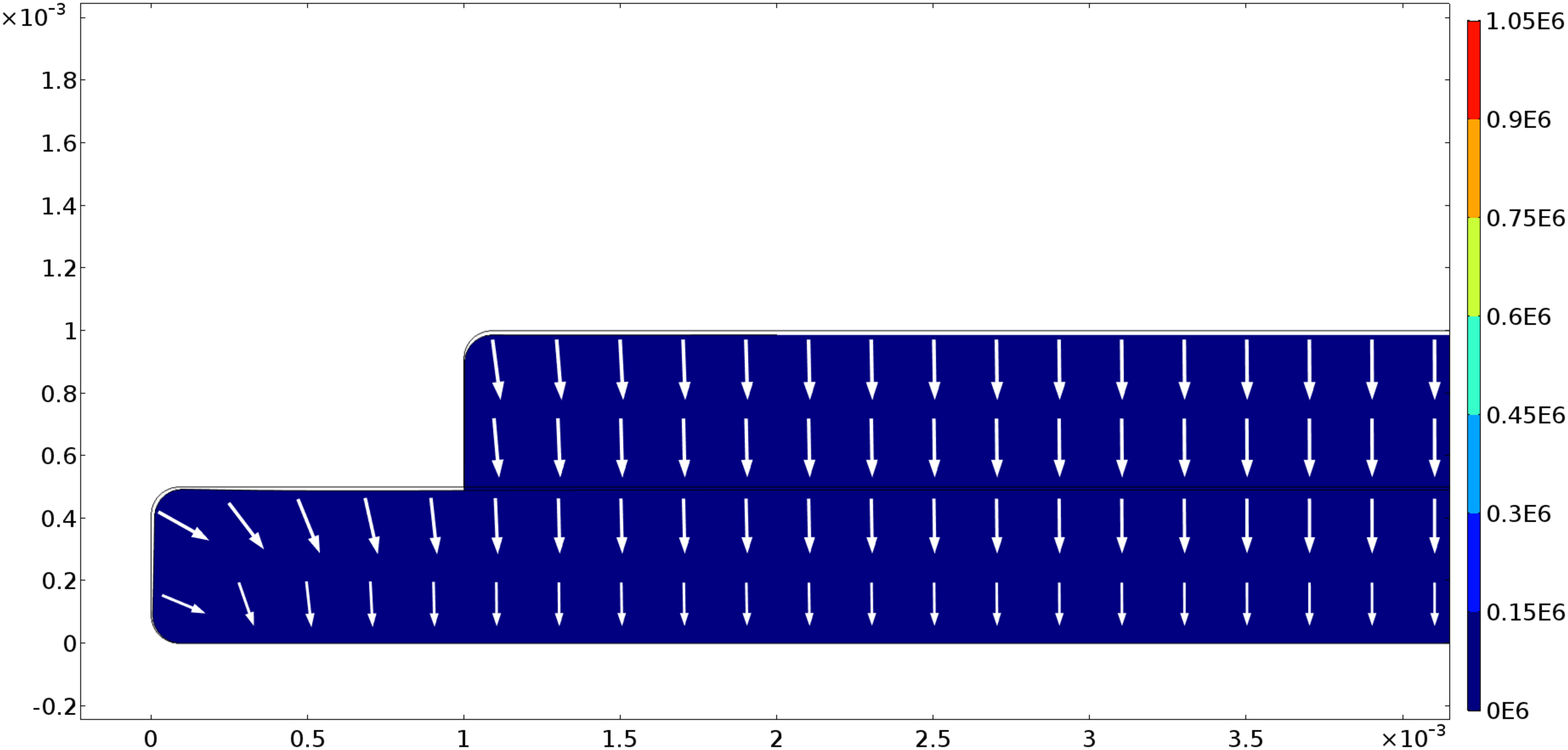}
                \caption{\scriptsize Distribution of the norm of deviatoric part of stress (Pa), $\mathrm{K}_2$, at 2 second. The value remains insignificant due to the re-heating of the material.}
                \label{eivc2}
        \end{subfigure}%
        \vspace{1cm}
\end{figure}
\begin{figure}\ContinuedFloat
\centering 
        \begin{subfigure}{1\textwidth}
                \includegraphics[width=14cm, height=6cm]{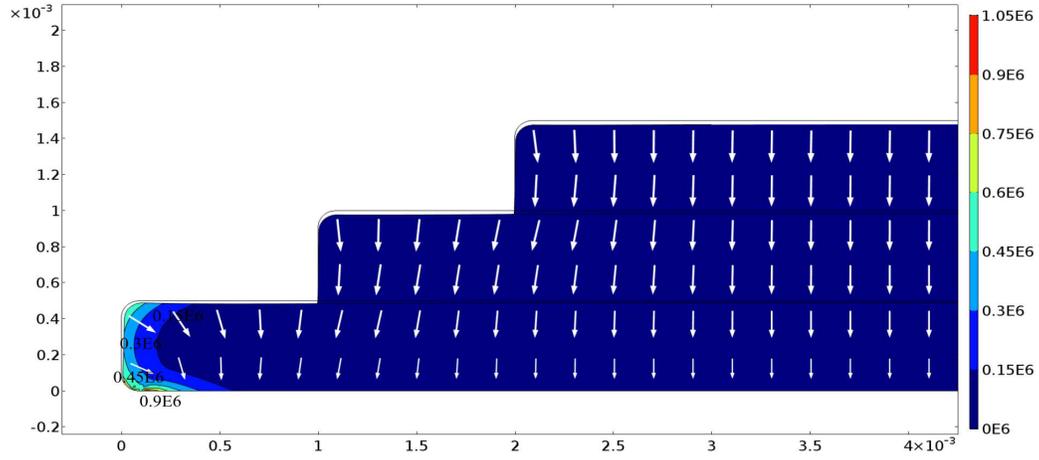}
                \caption{\scriptsize Distribution of the norm of deviatoric part of stress (Pa), $\mathrm{K}_2$, at 3 second. A significant amount of stress accumulation is observed in the region where the temperature values are close to $\theta_g$.}
                \label{eivc3}
        \end{subfigure}%
\vspace{1cm}
        \begin{subfigure}{1\textwidth}
                \includegraphics[width=14cm, height=6cm]{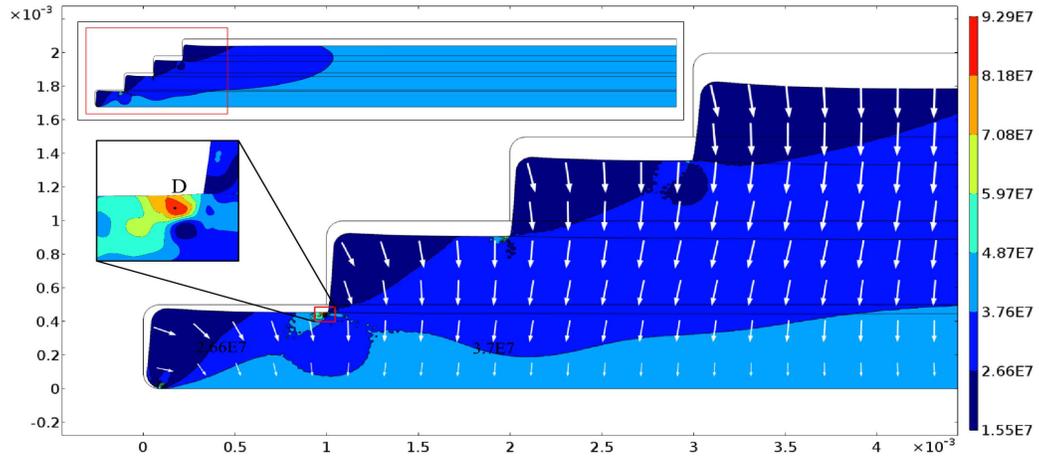}
                \caption{\scriptsize Distribution of the norm of deviatoric part of stress (Pa), $\mathrm{K}_2$, at 60 second. The inset shows a region having stress concentrations in the geometry. The point D inside this region is in a stress state given by: $\mathrm{T}_{xx}$ = 1.38E8, $\mathrm{T}_{yy}$ = 1.64E7, $\mathrm{T}_{zz}$ = 9.94E7, $\mathrm{T}_{xy}$ = 2.65E7. During service, a ductile failure is most likely to happen in the red zone containing the point D.}
                \label{eivc4}
        \end{subfigure}
        
        \caption{Banded contour plots of the norm of deviatoric part of stress (Pa), $\mathrm{K}_2$, visualised on the deformed configuration.}\label{exvol}
\end{figure}%
\begin{figure}[H]
\centering
\vspace{1cm}
        \begin{subfigure}{1\textwidth}
               \includegraphics[width=14cm, height=6cm]{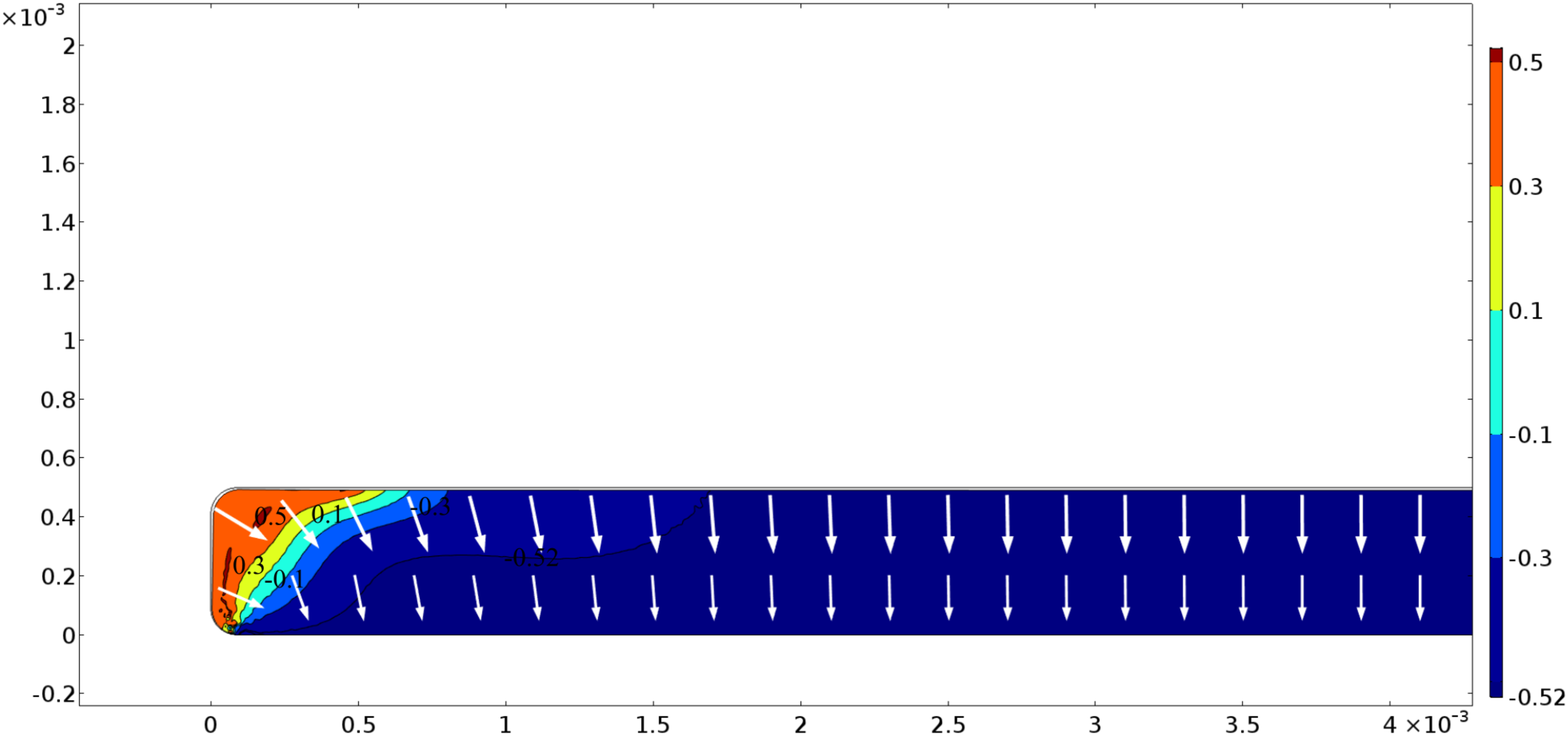}
                \caption{ \scriptsize Distribution of the modes of deformation ($\textrm{K}_3$) at 1 second.}
                \label{rvd1}
        \end{subfigure}%
        \vspace{2cm}
        \begin{subfigure}{1\textwidth}
                \includegraphics[width=14cm, height=6cm]{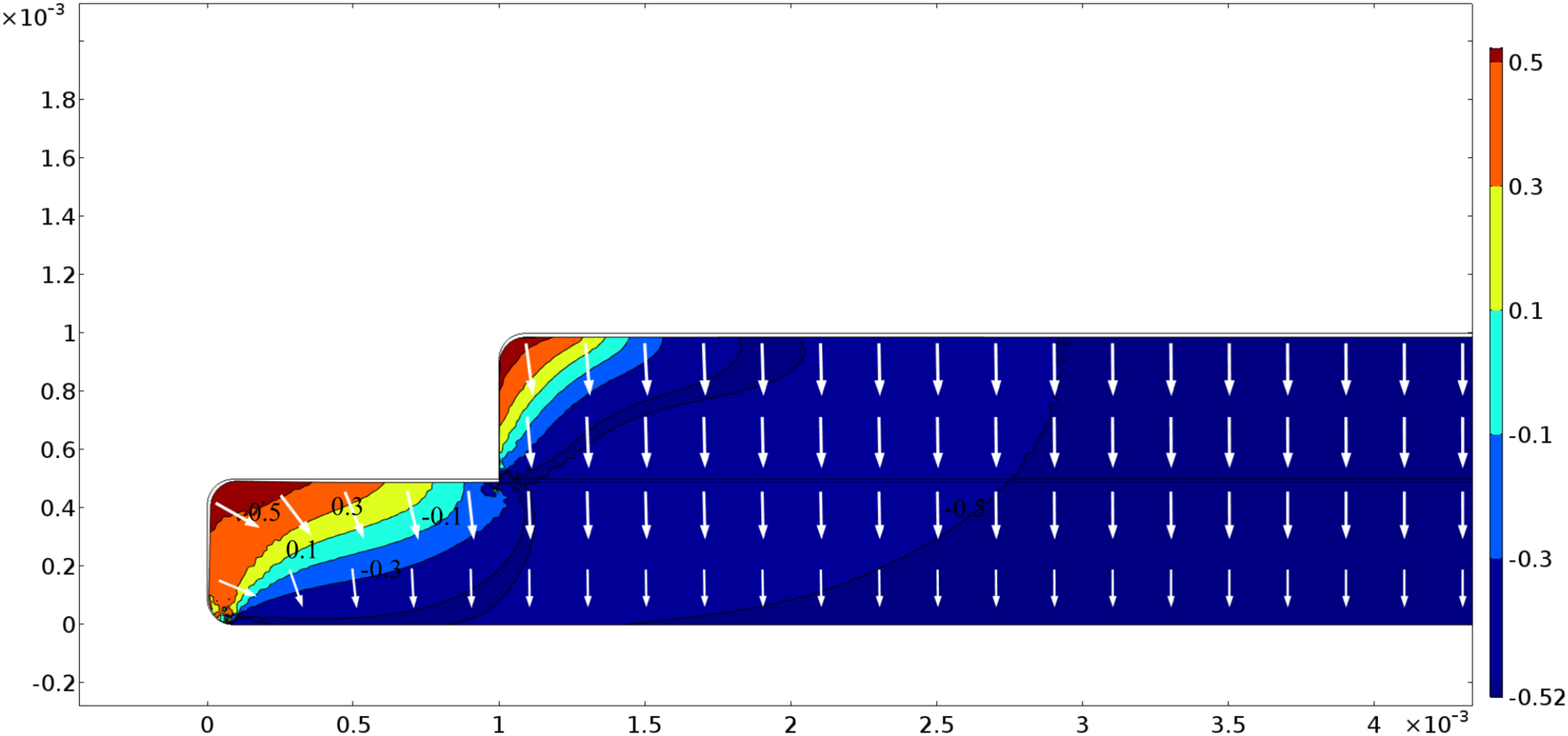}
                \caption{\scriptsize Distribution of the modes of deformation ($\textrm{K}_3$) at 2 second.}
                \label{rvd2}
        \end{subfigure}%
        \vspace{1cm}
\end{figure}
\begin{figure}\ContinuedFloat
\centering
        \begin{subfigure}{1\textwidth}
                \includegraphics[width=14cm, height=6cm]{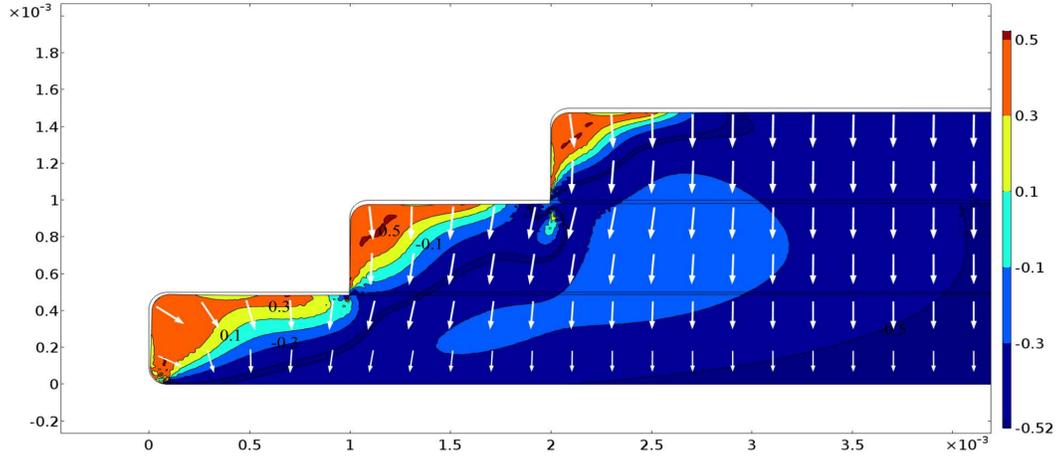}
                \caption{\scriptsize  Distribution of the modes of deformation ($\textrm{K}_3$) at 3 second.}
                \label{rvd3}
        \end{subfigure}%
\vspace{1cm}
        \begin{subfigure}{1\textwidth}
                \includegraphics[width=14cm, height=6cm]{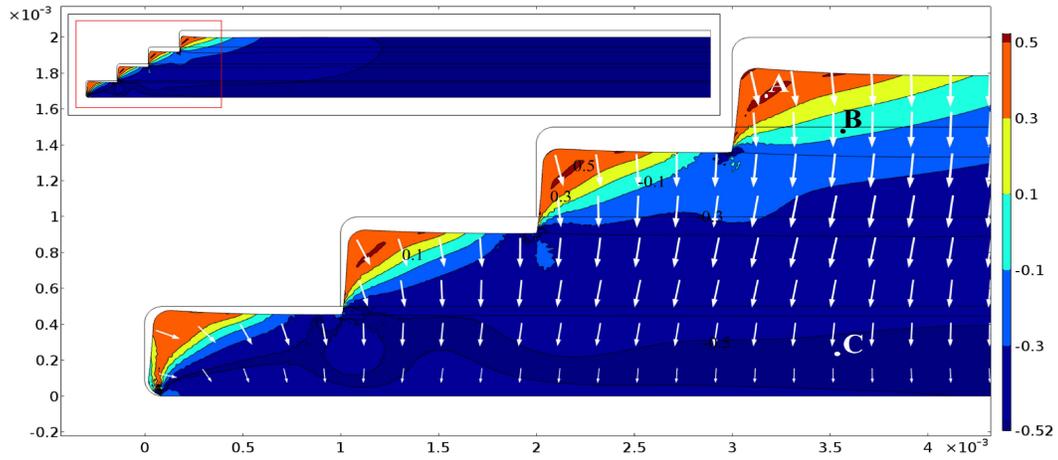}
                \caption{\scriptsize Distribution of the modes of deformation ($\textrm{K}_3$) at 60 second. Point A is close to uniaxial tension ($\mathrm{K}_3\approx0.52$), point B is close to pure shear ($\mathrm{K}_3\approx0$) and point C is close to equibiaxial tension ($\mathrm{K}_3\approx-0.52$).}
                \label{rvd4}
        \end{subfigure}
        \caption{Banded contour plots of the modes of deformation, ($\mathrm{\textrm{K}_3}$ invariant of the stress tensor), visualised on the deformed configuration. The dark red zones at the stepped end is close to uniaxial tension and dark blue zones at the core of the geometry is close to equibiaxial tension throughout the process.}\label{volde}
\end{figure}
 \begin{figure}[H]
\centering
\vspace{1cm}
        \begin{subfigure}{1\textwidth}
                \includegraphics[width=14cm, height=6cm]{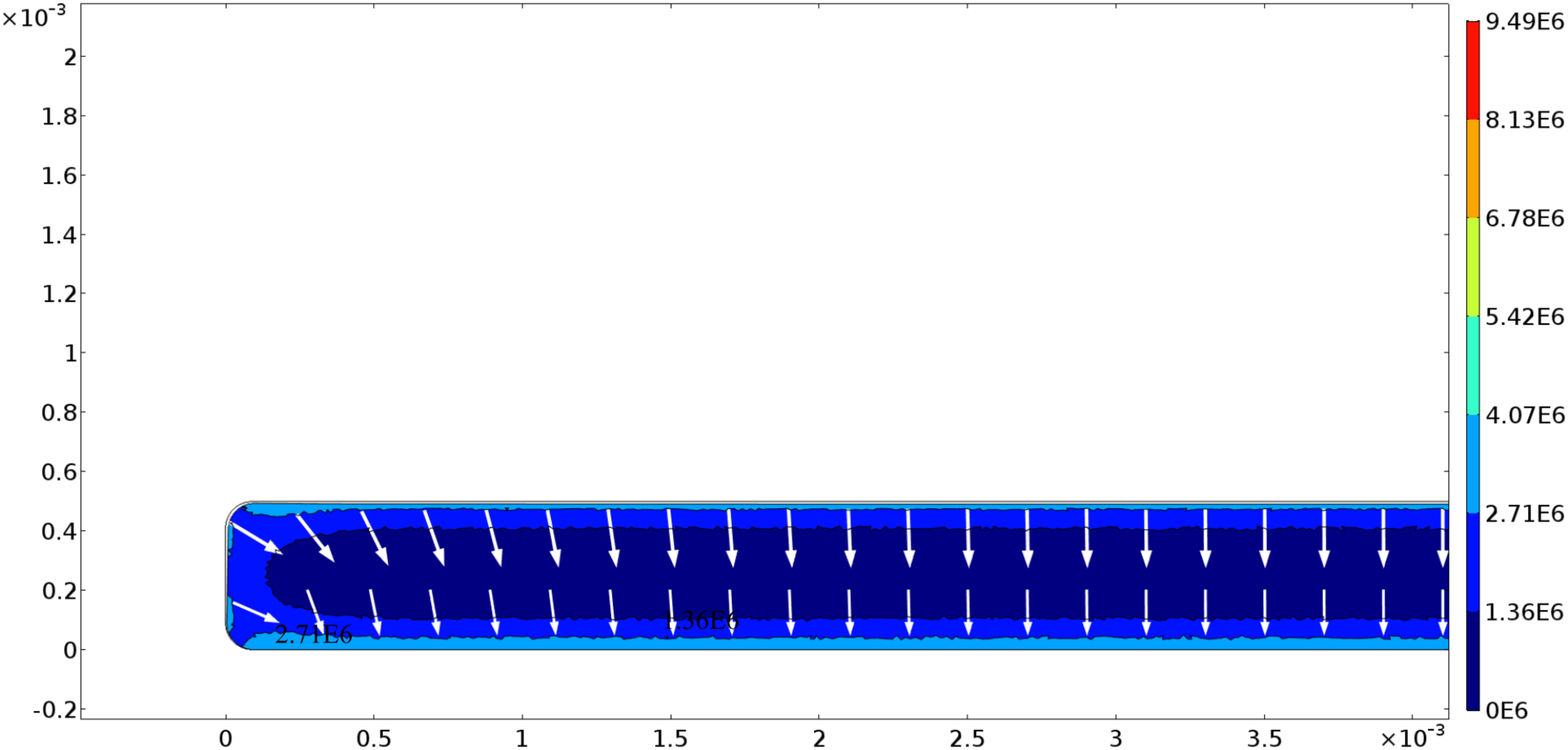}
                \caption{ \scriptsize Distribution of the total rate of entropy production ($\xi^{tot}_m$) at 1 second.}
                \label{ZETA1}
        \end{subfigure}%
\vspace{3cm}
        \begin{subfigure}{1\textwidth}
                \includegraphics[width=14cm, height=6cm]{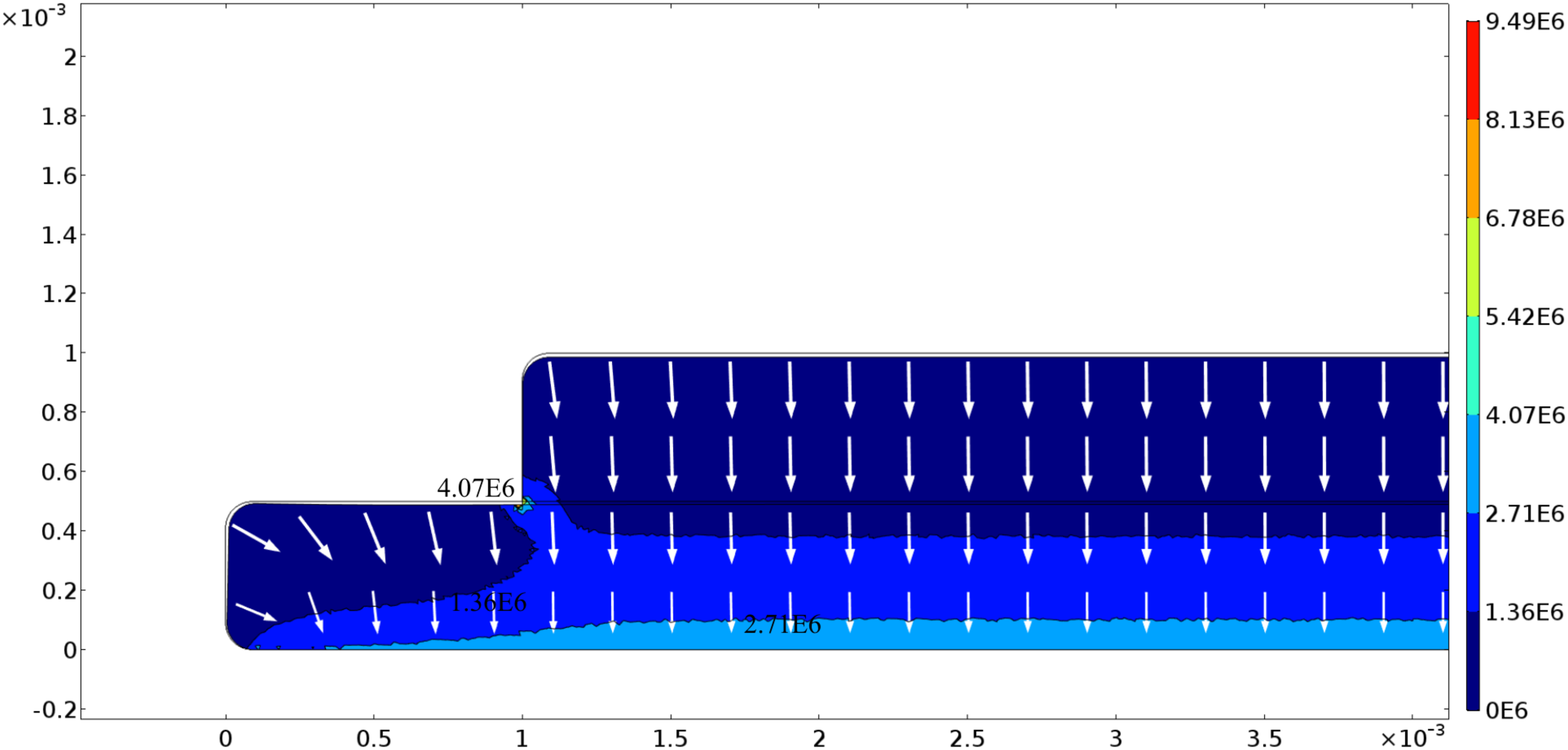}
                \caption{\scriptsize Distribution of the total rate of entropy production ($\xi^{tot}_m$) at 2 second.}
                \label{ZETA2}
        \end{subfigure}%
        \vspace{1cm}
\end{figure}
\begin{figure}\ContinuedFloat
\centering 
        \begin{subfigure}{1\textwidth}
                \includegraphics[width=14cm, height=6cm]{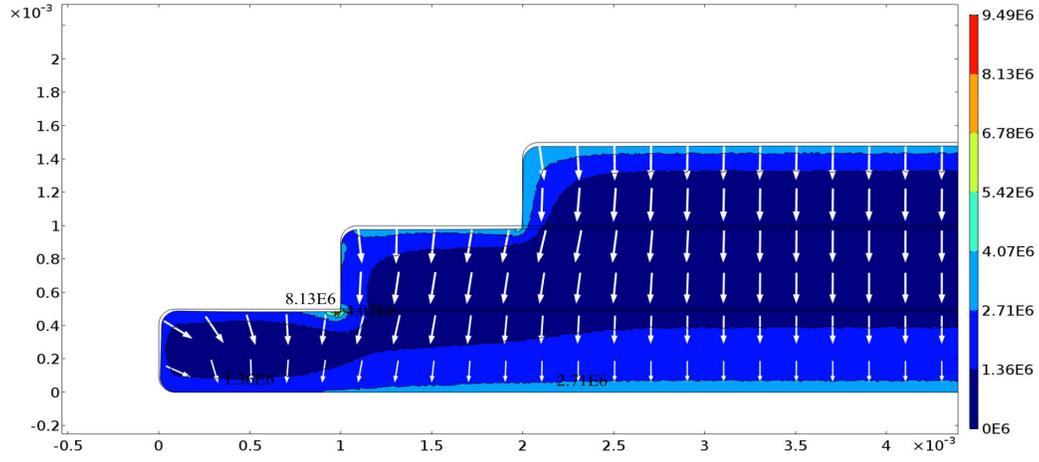}
                \caption{\scriptsize Distribution of the total rate of entropy production ($\xi^{tot}_m$) at 3 second.}
                \label{ZETA3}
        \end{subfigure}%
\vspace{2cm}
         \begin{subfigure}{1\textwidth}
            \includegraphics[width=14cm, height=6cm]{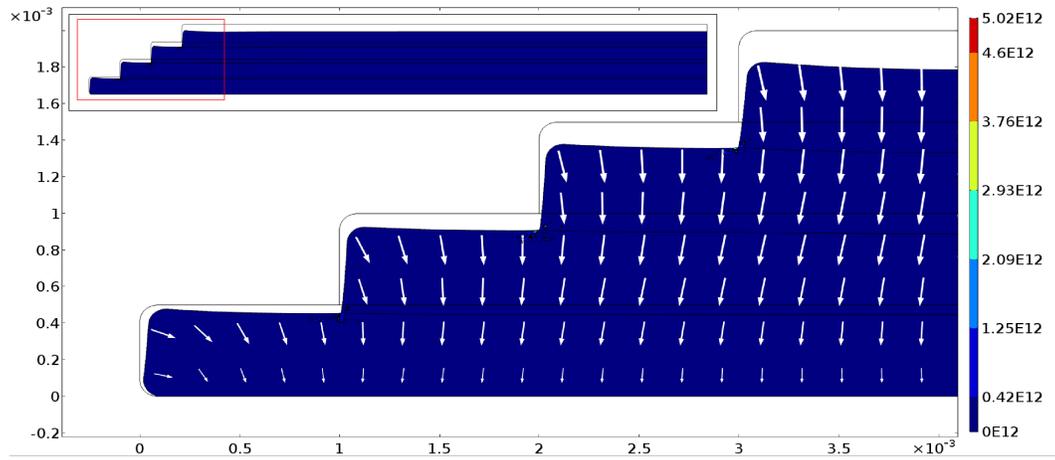}
                \caption{\scriptsize Distribution of the total rate of entropy production ($\xi^{tot}_m$) at 60 second.}
                \label{ZETA4}
        \end{subfigure}
        
        \caption{Banded contour plots of the total rate of entropy production ($\xi^{tot}_m$) visualised on the deformed configuration. }\label{ZETA}
\end{figure}%
 \noindent strikingly apparent. The reference configuration of the layers have been superimposed on all the plots, and it indicates the final volume to  be approximately $9\%$ less than the initial volume (refer to Figs.\ref{t4}-\ref{ZETA4}). The arrows superimposed on the plots represent the local displacement vector during the process.\\
 \indent As the material cools from the edges of the geometry, it is being pulled in, i.e., the $u_x$ components of the displacement field are positive and $u_y$ components are negative (the orientation of the basis are given in Fig.\ref{ri}). Towards the core, the $u_x$ components become negligible and only the $u_y$ components survive. However, when the existing layers are re-heated each time a new layer is laid, the core of the geometry starts expanding due to the inflow of heat, which leads to negative $u_x$ components and positive $u_y$ components at the core, although, the net effect of $u_y$ is in the negative direction throughout the material. Thus, the resultant displacement vectors near the corners would be directed downwards at an acute angle (clockwise) with respect to the $x$-direction, whereas the resultant displacement vectors at the core would be initially directed downwards (refer to Figs.\ref{t1}-\ref{ZETA1}), and later  due to re-heating, as time proceeds it would be directed downwards at an obtuse angle (clockwise) with respect to the $x$-direction (refer to Figs.\ref{t4}-\ref{ZETA4}).\\
 \indent The thermal and mechanical interactions lead to a complicated distribution of the residual stresses (refer to Fig.\ref{VOL} and Fig. \ref{exvol}) and consequently the modes of deformation (refer to Fig.\ref{volde}). The complexity is profound in the vicinity of the stepped geometry, whereas towards the opposite end, the distribution becomes one dimensional, i.e., in the $y$-direction. The magnitude of the stresses are negligible during the initial time steps (refer to Figs.\ref{VOL1}-\ref{VOL3} and Figs.\ref{eivc1}-\ref{eivc3}) due to the relatively lower volume shrikange. Re-heating of the layers, each time continuity is established, further adds to the relaxation of stresses.  Magnitude of the stresses shoot up drastically when the temperature falls below $\theta_g$, which is expected since the viscosity of the polymer increases sharply at $\theta_g$, and finally, the stresses freeze when the temperature is sufficiently below $\theta_g$ (refer to Fig.\ref{t4}, Fig.\ref{VOL4} and Fig.\ref{eivc4}). Another reason for a large magnitude of stresses being induced in the material (refer to Fig.\ref{VOL4} and Fig.\ref{eivc4}), is due to the high rate of cooling (refer to Table\ref{Table1}), which does not give enough time for the material to undergo stress relaxation.\\
\indent To understand the distribution of the modes, we consider a shear superposed unequi-triaxial stress state of the material points. We mark points A, B and C on three different zones corresponding to three different modes of deformation, as shown in the Figs.\ref{rvd4}. Note that the plane strain assumption constrains stress in the $z$-direction, i.e. $\mathrm{T}_{zz}$, to remain positive throughout the process. The stress state at point ``A" is given by: $\mathrm{T}_{xx}$=0.5MPa, $\mathrm{T}_{yy}$=0.73MPa, $\mathrm{T}_{zz}$=27.7MPa and $\mathrm{T}_{xy}$=-0.19MPa. The magnitude of the $\mathrm{T}_{zz}$ component is two orders higher than $\mathrm{T}_{xx}$, $\mathrm{T}_{yy}$ and $\mathrm{T}_{xy}$ components. Therefore, we can conclude that the mode of deformation of the material point ``A" is close to uniaxial tension in the z-direction with $\mathrm{K}_3\approx0.52$. Similarly, the mode of deformation of other points lying in the same zone will be close to uniaxial tension in the $z$-direction.\\
\indent Next, the stress state at point ``B" is given by: $\mathrm{T}_{xx}$=16.6MPa, $\mathrm{T}_{yy}$=-0.186MPa, $\mathrm{T}_{zz}$=34.8MPa and $\mathrm{T}_{xy}$=5.13MPa. The $\mathrm{T}_{xy}$ component is quite significant, and therefore, the zone in which point ``B" lies is close to a pure shear mode of deformation with $\mathrm{K}_3\approx0$. Finally, the stress state at point ``C" is given by: $\mathrm{T}_{xx}$=45.2MPa, $\mathrm{T}_{yy}$=-1.49MPa, $\mathrm{T}_{zz}$=47.5MPa and $\mathrm{T}_{xy}$=9.56MPa. $\mathrm{T}_{xx}$ and $\mathrm{T}_{zz}$ components are approximately same and positive. Also, they are one order higher than that of $\mathrm{T}_{yy}$ (which is negative) and $\mathrm{T}_{xy}$. Therefore, the mode of deformation at point ``C" is close to equibiaxial tension, and hence, the zone in which point ``C" lies has
$\mathrm{K}_3\approx-0.52$. Subsequently, we conclude that, the material at the top corners of all the layers are close to uniaxial tension, while the material at the core is close to equibiaxial tension. \\
 \indent The ratio of mean normal stress ($\frac{\mathrm{K}_1}{\sqrt{3}}$) to the norm of the deviatoric stresses ($\mathrm{K}_2$) is much less than 1 at the top corners of the layers and close to 1 at the core. Therefore, the corners are dominated by the distortional stresses, whereas, the core is dominated by the mean normal stress, implying that, dimensional instability is more at the corners than at the core. This is further supported by the bulging and the warping observed in the neighbourhood of the stepping at the final stages of the simulation (refer to  Figs.\ref{t4}-\ref{ZETA4}). Warping can also be understood based on the direction of the components of the displacement fields at the top corners, where, $u_x$ is positive and $u_y$ is negative. Therefore, the material at the topmost corner will get pushed out, leading to warping.\\
 \indent During the entire process, the total rate of entropy production has to be maintained positive. This needs to be checked a posteriori at every time step of the process. The plots of total rate of entropy production at some chosen time steps have been given in Figs.\ref{ZETA1}-\ref{ZETA4}, and the values remain positive as expected.
 \subsubsection{Stress concentrations and delamination}
 \begin{figure}[H]
\centering
                \includegraphics[width=14cm, height=6.5cm]{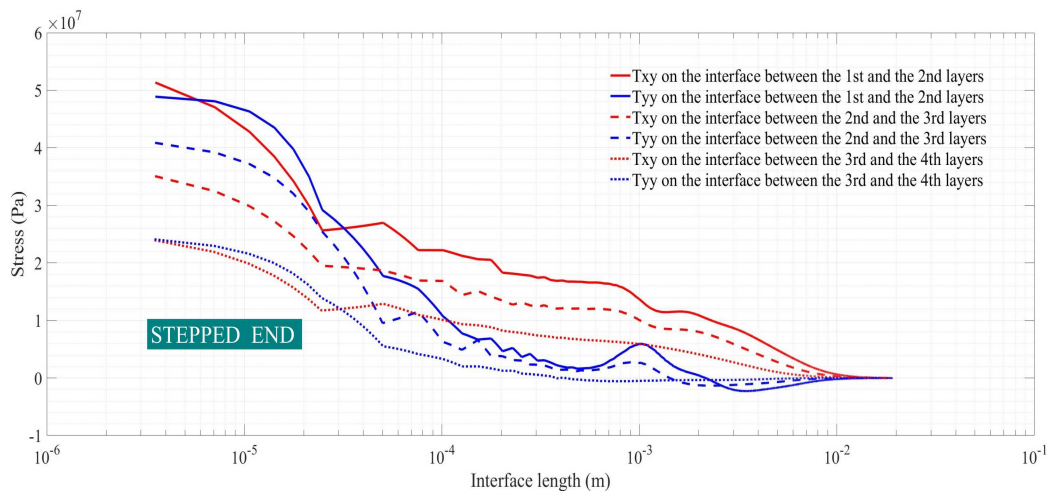}
        \caption{The stress components ($\mathrm{T}_{xy}$ and $\mathrm{T}_{yy}$) that are responsible for delamination (combination of mode I and mode II) of the layers, visualised along the length of all the three interfaces (semi-log plot), after the melt has solidified (t = 60 second). }\label{delamination}
\end{figure}%
 \indent In a practical setting, the interfaces between the layers have discontinuities, and therefore, the inter-layer strength is usually the weakest, which often causes delamination of the layers for a component fabricated by FDM. Even though, in the current analysis, the interfaces are perfect, we assume that the layers have a ``tendency" to delaminate. Fig.\ref{delamination} helps us visualise the stress components responsible for delamination along the length of each of the interfaces. Only $\mathrm{T}_{yy}$ and $\mathrm{T}_{xy}$ components of the stress tensor cause delamination. $\mathrm{T}_{yy}$ component tends to pull the layers apart (mode I delamination), whereas, $\mathrm{T}_{xy}$ component tends to slide the layers in the plane of the interface (mode II delamination). It is observed that, $\mathrm{T}_{yy}$ component becomes compressive in the region that is sufficiently away from the stepped end. Therefore, in this region, delamination can occur only due to the $\mathrm{T}_{xy}$ component (mode II delamination). The  tendency to delaminate is the highest for the interface between the first and the second layer.\\
 \indent The effect of mode I and mode II delamination tendencies, and the traction free condition on the free edges, lead to the formation of stress concentrations in the neighbourhood of the stepping (refer to Fig.\ref{eivc4}). If delamination occurs, and the interface breaks apart near the stepping, the stress concentrations would be relieved. The stress components are the highest (order of seven and eight) in the region of the stress concentration of the first layer. Therefore, yielding is most likely to occur in this region and lead to a ductile failure termed as ``intra-layer fracture". Usually, inter-layer fracture due to delamination and other discontinuities is the most common type of failure, compared to the intra-layer fracture caused by stress concentrations (see Macedo et. al. \cite{Quelho}). Therefore, a component fabricated by FDM, when in service, will most probably fail due to the delaminated interfaces termed as``inter-layer fracture".
\section{Conclusion}
A thermodynamic framework was developed to determine the differential volumetric shrinkage caused by drastic temperature gradients, the ensuing residual stresses and consequently the dimensional instability of the geometry, by building on an existing theory which required the definition of two primitives: Helmholtz free energy and the rate of entropy production. The theory is most suitable for processes that use amorphous polymers or polymers which do not crystallize much. The efficacy of the constitutive relation was verified by considering a prototypical FDM process in which four layers of polystyrene melt were laid, such that, plane strain conditions could be enforced.\\
\indent As soon as the layers were laid, it underwent a complex thermo-mechanical process which included rapid cooling, reheating, solidification at the glass transition temperature ($\theta_g$) and redistribution of the residual stresses. Although, a high value of the shear modulus caused stresses in the range of kilo Pascals in the melt phase, it is quite insignificant as compared to the stresses in the solid phase. This conforms with the reasoning put forward by  Xinhua et. al. \cite{Xinhua}, Wang et. al. \cite{Wang}, Park et. al. \cite{KeunPark} and Macedo et. al. \cite{Quelho}, who have confined the analysis to the solidified polymer since the stresses in the melt phase can be assumed to be negligible. The mean normal stress distribution at the final time step (refer to Fig.\ref{VOL4}) seems to follow a similar trend, in comparison to the mean normal stress distribution reported by Xia and Lu et. al. \cite{FDM2} (refer to Fig.8 in \cite{FDM2}).\\
\indent The plane strain approximation caused the outer corners of the layers to be close to uniaxial tension and the core to be close to equibiaxial tension. During service, there is a high probability for a pre-existing crack at the core to open up in the $x$-$z$ plane due to the equibiaxial tension. Further, in a practical setting, analogous to the current analysis, if the interfaces are highly porous, a large amount of inter-layer fracture (a combination of mode I and mode II) is expected at the stepped ends as compared to intra-layer fractures caused by stress concentrations. Another major dimensional instability that occurred was the warping of the geometry.  Warping can be reduced by making the corners in the stepping smoother. More smooth the corners are, less the degree of warping. The possibility of delamination, warping and the magnitudes of the stress concentrations, can be reduced by choosing the right combination of rate of cooling, substrate temperature, re-heating and the geometry for the PS layers, which reduces the magnitude of the ``freezed in" stresses (see Macedo \cite{Quelho}).
\clearpage
\newpage

\section*{References}
\bibliographystyle{elsarticle-num}
 \bibliography{REFERENCES}
% \newpage
% \appendix
%\appendixpage
%\section{Rate form of stress}
\end{document}